\documentclass
[pra,aps,twocolumn,showpacs,lengthcheck,preprintnumbers,superscriptaddress]{revtex4-1}%
\usepackage{amsfonts}
\usepackage{amsmath}
\usepackage{amssymb}
\usepackage{graphicx}
\usepackage{color}
\usepackage[normalem]{ulem}
\newcommand{\floor}[1]{\lfloor #1 \rfloor}
\usepackage[unicode]{hyperref}
\hypersetup{
   unicode=true,          
   plainpages=false,
   colorlinks=true,       
   citecolor=blue,        
}

\begin{document}

\author{A. Mu\~{n}oz Mateo}
\affiliation{
Centre for
Theoretical Chemistry and Physics, New Zealand Institute for Advanced Study,
Massey University, Private Bag 102904 NSMC, Auckland 0745, New Zealand
}
\affiliation{Dodd-Walls Centre for Photonic and Quantum Technologies, New Zealand}
\affiliation{Departament de F\'{i}sica Qu\`antica i Astrof\'isica, 
Universitat de Barcelona, Mart\'i i Franqu\`es 1, E--08028 Barcelona, 
Spain }
\author{V. Delgado}
\affiliation{Departamento de F\'{i}sica, Universidad de La Laguna, E--38205, 
La Laguna, Spain}
\author{M. Guilleumas}
\affiliation{Departament de F\'{i}sica Qu\`antica i Astrof\'isica, 
Universitat de Barcelona, Mart\'i i Franqu\`es 1, E--08028 Barcelona, 
Spain }
\affiliation{Institut de Ci\`encies del Cosmos, Universitat de 
Barcelona, ICC-UB, E--08028 Barcelona, Spain}
\author{R. Mayol}
\affiliation{Departament de F\'{i}sica Qu\`antica i Astrof\'isica, 
Universitat de Barcelona, Mart\'i i Franqu\`es 1, E--08028 Barcelona, 
Spain }
\affiliation{Institut de Ci\`encies del Cosmos, Universitat de 
Barcelona, ICC-UB, E--08028 Barcelona, Spain}
\author{J. Brand}
\affiliation{
Centre for
Theoretical Chemistry and Physics, New Zealand Institute for Advanced Study,
Massey University, Private Bag 102904 NSMC, Auckland 0745, New Zealand
}
\affiliation{Dodd-Walls Centre for Photonic and Quantum Technologies, New Zealand}

\title{Nonlinear waves of Bose-Einstein condensates in rotating ring lattice potentials}

\begin{abstract} 
We analyze the dynamics of Bose-Einstein condensates loaded in 
rotating ring lattices made of a few sites, and show how rotation maps the 
states found in this finite system into those belonging to a static infinite 
lattice. Ring currents and soliton states in the absence of a lattice find their
continuation in the presence of the lattice as nonlinear Bloch waves and 
soliton-like states connecting them.
Both bright gap solitons and 
dark-soliton trains are shown to connect continuously to linear solutions. 
The existence of adiabatic paths upon varying rotation frequency between states with quantized
supercurrents suggests highly controllable
methods for the experimental generation of persistent currents.

\end{abstract}


\maketitle

\section{Introduction}
The experimental achievement of preparing, manipulating, and probing ultracold
gases in periodic 
potentials has made it possible to observe elusive phenomena that typically 
belong to the realm of solid state physics. Ultracold 
atoms can quantum-simulate strongly-interacting many-body 
systems \cite{Bloch2012} as well as dynamical processes in nonlinear media  
\cite{Morsch2006}. 
In the regime of large particle number per lattice site, the superfluid dynamics
exhibited by Bose-Einstein condensates (BECs) loaded in (approximately) 
periodic potentials created by optical interference phenomena
(optical lattices) provides an excellent sample of nonlinear phenomena. These 
range from the realization of atomic Bloch waves \cite{Dahan1996} that mimic 
the motion of electrons in metals,  to the generation of solitonic 
structures \cite{Eiermann2004}, which are the atomic 
equivalent of the nonlinear pulses currently used to carry information through 
optical fibers \cite{Mollenauer2006}.

The early studies of BECs in periodic potentials were mainly focused on
linear lattices, including one- (1D), two- (2D) and three-dimensional (3D)
lattices \cite{Dahan1996,Burger2001,Greiner2001,Greiner2002}. In these 
settings, the superfluid flow has been demonstrated 
to support many types of nonlinear waves  
\cite{Louis2003,Mateo2011,Mateo2014}.  The stability of these is intimately
connected with the possibility for breakdown of superfluid flow and has been 
studied extensively \cite{Burger2001,Wu2001,Wu2003,Kramer2003}.
In the last years, with the advent 
of new experimental techniques capable of imprinting lattices along a ring 
\cite{Franke-Arnold07,Henderson2009,Gauthier2016}, there has been an 
increasing interest in these configurations. They fulfill the Born-von Karman 
boundary conditions that are presupposed in modeling crystals \cite{Kittel1996}. 
Most of the previous theoretical studies on this subject have
started from discrete lattice models (see Refs.
\cite{Amico2014,Aghamalyan2015,Kolavr2015,Arwas2015,
Gallemi2015,Gallemi2016,Cataldo2011,Paraoanu2003} and references therein), 
while nonlinear waves in continuous models with periodic potentials has
received comparatively less attention 
\cite{Wu2000,Bronski2001,Diakonov2002,Machholm2003,Mueller2002}.

The ring geometry introduces interesting features on its own, mostly because of 
its suitability for the study of metastable persistent currents  
\cite{Ryu2007,Moulder2012}. In spite of the fact that persistent currents can also 
be found at a microscopic scale in normal metals 
\cite{Bleszynski-Jayich2009,Bluhm2009}, metastable persistent currents 
are one of the hallmarks of superfluidity
\cite{Reppy1964,Ramanathan2011,Beattie2013,Cherny2012} and superconductivity 
\cite{Deaver1961,Matveev2002}.
Their generation and decay through phase-slip events in toroidal BECs stirred 
by optically induced weak links are currently the object of intensive research,
since these systems have opened up a promising way of advancing towards the 
realization of atomtronic devices \cite{Amico2005,Cominotti2014,Eckel2014,Mateo2015}.

The present work aims to delve into the analysis of repulsively interacting 
BECs loaded in continuous ring lattices made of a small number of sites. By driving the 
lattice into angular rotation, the system exhibits the interplay of the rotating 
periodic potential and the intrinsic periodicity of the induced ring currents. 
In this situation, there also exists a relevant connection between 
angular quasimomentum and mean angular momentum. 
Previous analysis from nonlinear optics \cite{Kartashov08} has demonstrated 
that the ring lattice settings introduce significant changes in the 
properties of the associated nonlinear waves. 
However, such studies have only considered a particular type of nonlinear waves 
present in systems with a relatively large number of lattice sites, hence sharing 
many similarities with the limiting case of the infinite linear lattice. On 
the contrary, our approach starts by considering configurations containing a 
small number of wells (1, 2, 3, 4, \dots) from the perspective of a periodic array. 
We study how the    superfluid properties  change as the number of 
wells and the nonlinear interaction strength are increased, approaching finally the 
well known limit of an infinite lattice \cite{Wu2003}. This approach allows us 
to show in a unified and simple way the emergence of different types of 
nonlinear waves, namely Bloch 
waves, bright gap solitons and dark solitons. Although these types of 
genuinely nonlinear states have been extensively studied in infinite lattices, 
they do not share a common theoretical framework and have so far been lacking  a
proper characterization in the case of finite, short lattices. Furthermore, we 
demonstrate the continuation of the nonlinear states into the non-interacting, 
linear regime, and discuss adiabatic pathways through parameter space to change 
the current and winding number, extending the work of Refs.\ 
\cite{Kanamoto2009,Fialko2012}. Our results contribute towards a cohesive 
picture of the physical phenomena in optical lattices. 

\section{System: the ring lattice}

We consider a Bose-Einstein condensate confined to quasi-1D by a ring 
trap of mean radius 
$R$ at zero temperature and in the presence of a periodic potential along the
azimuthal coordinate $\theta$  that rotates with angular frequency
$\Omega$. By assuring that the number of particles per potential well is large 
enough for a mean-field model to be safely applied, the dynamics
of the order parameter $\psi(\theta,t)$ in the co-rotating frame of reference 
follows the 1D Gross-Pitaevskii equation
\begin{equation}
  i \hbar \,{\partial_t}\psi= \left\lbrace
\frac{\left(-i{\hbar}\,{\partial_\theta}-m\Omega 
R^2\right)^2}{2m\,R^2}
  + V(\theta) + g |\psi|^2
  \right\rbrace\psi \, ,
  \label{eq:1DGPt}
\end{equation}
where $V(\theta)=V_0 \sin^2({\pi R \theta}/{d})$  is the 
periodic potential introduced by an optical lattice with lattice constant
$d$ and well depth $V_0$. The  inter-particle contact interaction has strength
$g=2\hbar^2 a/m a_\perp^2$, determined by the $s$-wave scattering length $a$ 
and the size $a_\perp$ of a tight transverse confinement. The wave function 
$\psi$ is normalized to the number of particles in the ring  $N= R \int
d\theta\, |\psi|^2$. 

The ring geometry provides a wave-number quantum $k_R=1/R$ imposed 
by the finite size of the system, so that the smallest amount of  
kinetic energy available in the ring is $E_R/2$, where
$E_R=\hbar^2\,k_R^2/m=\hbar\, \Omega_R$. This fact turns out to be crucial 
for the comparison with results obtained in the infinite lattice, where $E_R=0$.
On the other hand, the periodic potential introduces the characteristic 
reciprocal lattice vector $k_L=2\pi/d$, which defines a maximum quasimomentum 
$\hbar k_L/2$ at the end of the Brillouin zone and a corresponding lattice 
energy, or recoil  energy, $E_L=\hbar^2 \pi^2/2\,m \,d^2$. Both, 
$k_R$ and $k_L$, are related by the commensurable ratio $M=k_L/k_R=2\pi R/d$, 
which gives the number of lattice sites in the ring. 

By writing Eq.~(\ref{eq:1DGPt}) in the ring energy units $E_R$, and searching 
for the stationary states $\psi(\theta,t)=\psi(\theta)\,e^{-i\mu t/\hbar}$, 
with chemical potential $\mu$, we obtain a stationary nonlinear Schr\"odinger
equation in dimensionless form
\begin{equation}
  \left\lbrace
\frac{1}{2}\left(-i\partial_\theta-\hat{\Omega} \right)^2
  + \hat{s} \sin^2\left(\frac{M}{2} \theta\right)+  
\hat{g} |\hat{\psi}|^2
  - \hat{\mu}\right\rbrace\hat{\psi} = 0\, ,
  \label{eq:1DGP}
\end{equation}
where $\hat{\mu}=\mu/E_R$, $\hat{\Omega}=\Omega/\Omega_R$, $\hat{g}=2a R/ 
a_\perp^2$, the rescaled wave function is $\hat{\psi}= \sqrt{R} \, 
\psi$, and the lattice depth $V_0$ is measured relative to the ring energy
$\hat{s}=V_0/E_R = s M^2/8$, with $s=V_0/E_L$. It is worth noticing that 
$s$, instead of $\hat{s}$, is the usual parameter for identifying the dynamical 
regimes in the presence of the lattice, from the shallow lattice 
regime ($s\lesssim 1$) up to the tight binding limit ($s\gg 1$) 
\cite{Louis2003,Kramer2003,Mateo2011,Mateo2014}. For later use, we also define 
an average interaction parameter per lattice site $\eta=\hat g N/M$.

From Eq.\ (\ref{eq:1DGPt}), the local conservation of particle number in the
rotating frame, $\partial_\tau |\hat{\psi}|^2 + \partial_\theta \hat J=0$ 
(with $\tau=t\,\Omega_R$), determines the current density 
\begin{align}
\hat J(\theta) = |{\hat\psi(\theta)}|^2\, \left(\hat p(\theta)- 
\hat \Omega\right)
 \label{eq:current}
\end{align}
where $\hat p = \partial_\theta \arg(\psi)$ is the local
canonical momentum density. 
The superfluid 
velocity is given by
 $\hat v=\hat J/ |{\hat\psi}|^2$.
The mean angular momentum per particle $L_z$ 
(along the direction perpendicular to the ring) follows from
\begin{align}
L_z=\frac{\hbar} {N} \oint d\theta \, |{\hat\psi}(\theta)|^2 \, \hat 
p(\theta).
\label{eq:Lz}
\end{align}

The linear excitation modes $[u(x,t),v(x,t)]$ with energy 
$\hat\omega=\hbar\omega/E_R$ around the stationary state $\hat\psi$ fulfill the 
Bogoliubov equations \cite{Pitaevskii2016}
\begin{align}
 (H_L+ 2\, \hat g \,|\hat\psi|^2 )\, u+  \, \hat g  \, \hat\psi^2 v= 
\hat\omega \, u 
\nonumber\\
 -(H_L^*+ 2\, \hat g \,|\hat\psi|^2 )\, v-\, \hat g \, (\hat\psi^*)^2 u= 
\hat\omega \, v 
\,,
\label{eq:bog}
\end{align}
where $H_L = (-i\partial_\theta-\hat{\Omega})^2/2  + \hat{s}\sin^2(M \theta/2) 
-\hat\mu$. 
The solutions to the Bogoliubov equations provide information about
the linear dynamical stability in the ring 
lattice. In particular the presence of complex eigenfrequencies Im$[\hat\omega]\neq0$ 
indicates that the stationary state is dynamically unstable and prone to 
decay, since the associated unstable modes, if excited,
grow exponentially $\propto \exp(\mathrm{Im}[\hat\omega]\,t)$.

It is instructive to start reviewing the eigenstates of the non-interacting 
($g=0$) ring lattice. At $\hat\Omega$=0, the stationary states have 
definite 
quasimomentum and play an equivalent role to that of winding number states, 
with definite angular momentum in the homogeneous ring. This equivalence still 
applies for nonzero $\hat\Omega$, by means of which the stationary states get 
their energies shifted inside the energy bands.

\subsection{Non-interacting regime}

According to the Bloch theorem, the eigenstates of the non-rotating
system are the Bloch waves $\psi_{n,k}(R\theta)=e^{ikR\theta}u_{n,k}(R\theta)$, 
where $\hbar k$ is the linear quasi- (or crystal) momentum that takes values 
within the first Brillouin zone $\hbar\, (-k_L/2, k_L/2]$. The function 
$u_{n,k}$ is periodic, and shares the period $d$ of the lattice 
\cite{Kittel1996}, such that $u_{n,k}(R\theta)=u_{n,k}(R\theta+d)$, where $n$ 
is the  
band index, so that the Bloch waves are single valued in the ring, i.e. 
$\psi_{n,k}(R\theta)=\psi_{n,k}(R(\theta+2\pi))$. Due to 
the finite size of the 
system, the quasi-momentum takes discrete values $k=q \,k_R= q/R$, with 
$q=0,\pm1,\dots, M/2$ being an integer number, and so the ring 
lattice bands exhibit a discrete structure. 

From now on we will denote the Bloch waves by the two quantum numbers 
$|n,q\rangle\equiv \psi_{n,q}(\theta)=e^{iq\theta}u_{n,q}(\theta)$, or simply 
$|q\rangle$ if the band is implicitly assumed.
There are just $M$ Bloch waves within the first Brillouin zone. In particular, 
when the lattice contains an even number of sites the crystal angular momentum 
reaches the edge 
of the Brillouin zone at $q=M/2$, while this limit is not reachable when the 
number of sites is odd, and the maximum $|q|$ are $\pm q_M$, where $\{ 
q_{M}=\floor{M/2}\}<M/2$ is the maximum integer less than $M/2$.

\begin{figure}[tb]
 \centering
  \includegraphics[width=\linewidth]{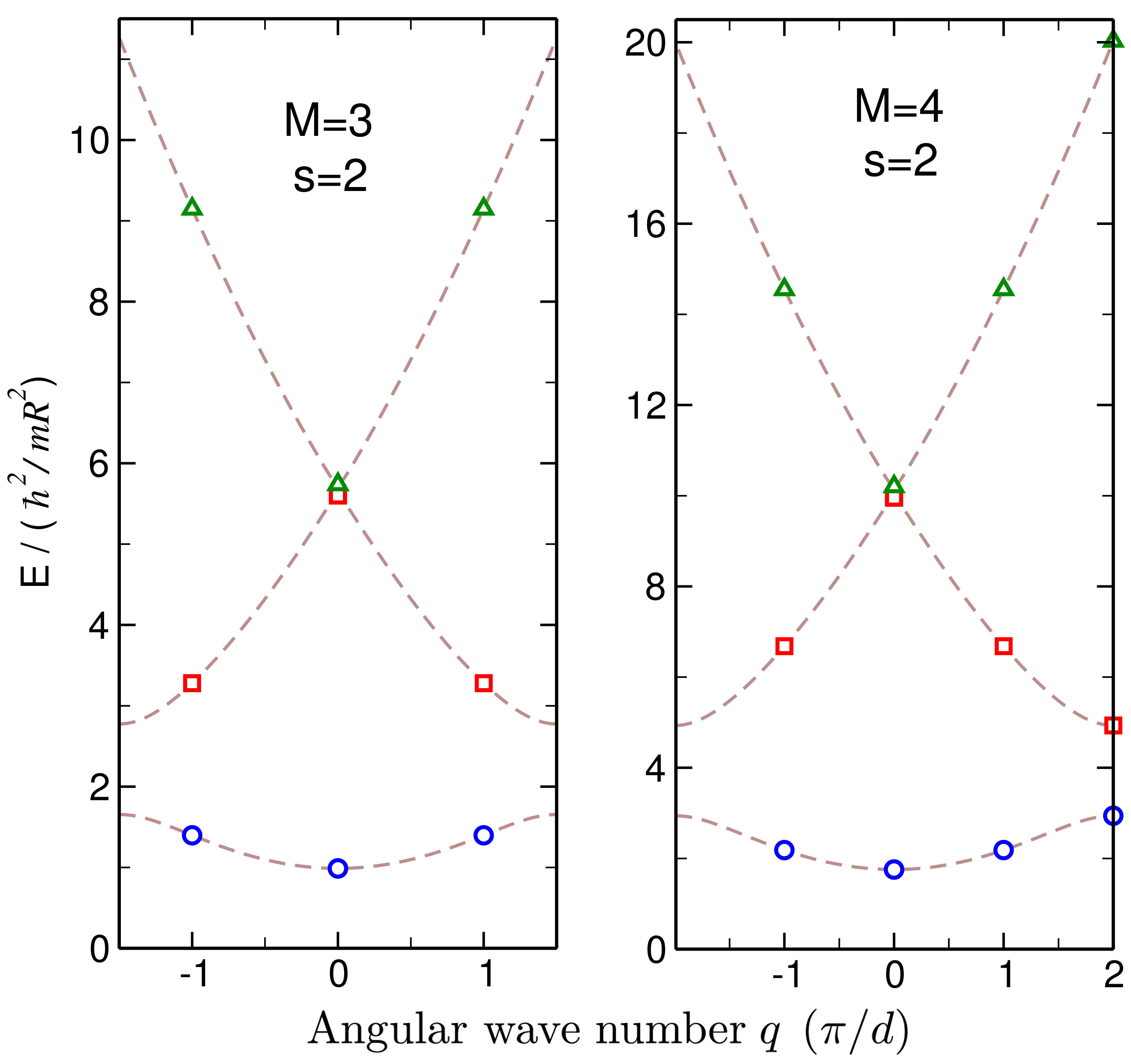}
 \caption{Energy of the linear Bloch states (symbols) in the three lowest 
energy bands $n=1, 2, 3$ of non-rotating ring lattices
with site numbers $M=3$ (left panel) and $M=4$ (right panel), sharing depth 
$s=V_0/E_L=2$ and radius $R$. The energy difference between the two systems 
is due to the different lattice spacing $d$. The continuous bands of the 
infinite lattices, with equal corresponding $d$, are represented by 
dashed lines.}
 \label{fig:M34band}
\end{figure}
For instance, let us consider two ring lattices, sharing radius $R$ and depth 
$V_0=2 E_L$, with $3$ and $4$ sites. The energy eigenvalues $E$ versus 
quasimomenum $q$ are represented in Fig.\ \ref{fig:M34band} (note that 
$E=\mu$ in this regime). There 
are three ($q=0, \pm1$) and four ($q=0, \pm1, 2$) Bloch states per band, 
respectively. While for $M=4$ there is a single maximum-quasimomentum
eigenstate ($q=2$) at the band edge, for $M=3$ there are two 
(energy-degenerate) maximum-quasimomentum eigenstates $q=\pm q_3=\pm 1$ that 
belong to each band interior.

\subsection{Energy degeneracies}
\label{sec:degeneracies}

\begin{figure}[tb]
 \centering
 \includegraphics[width=\linewidth]{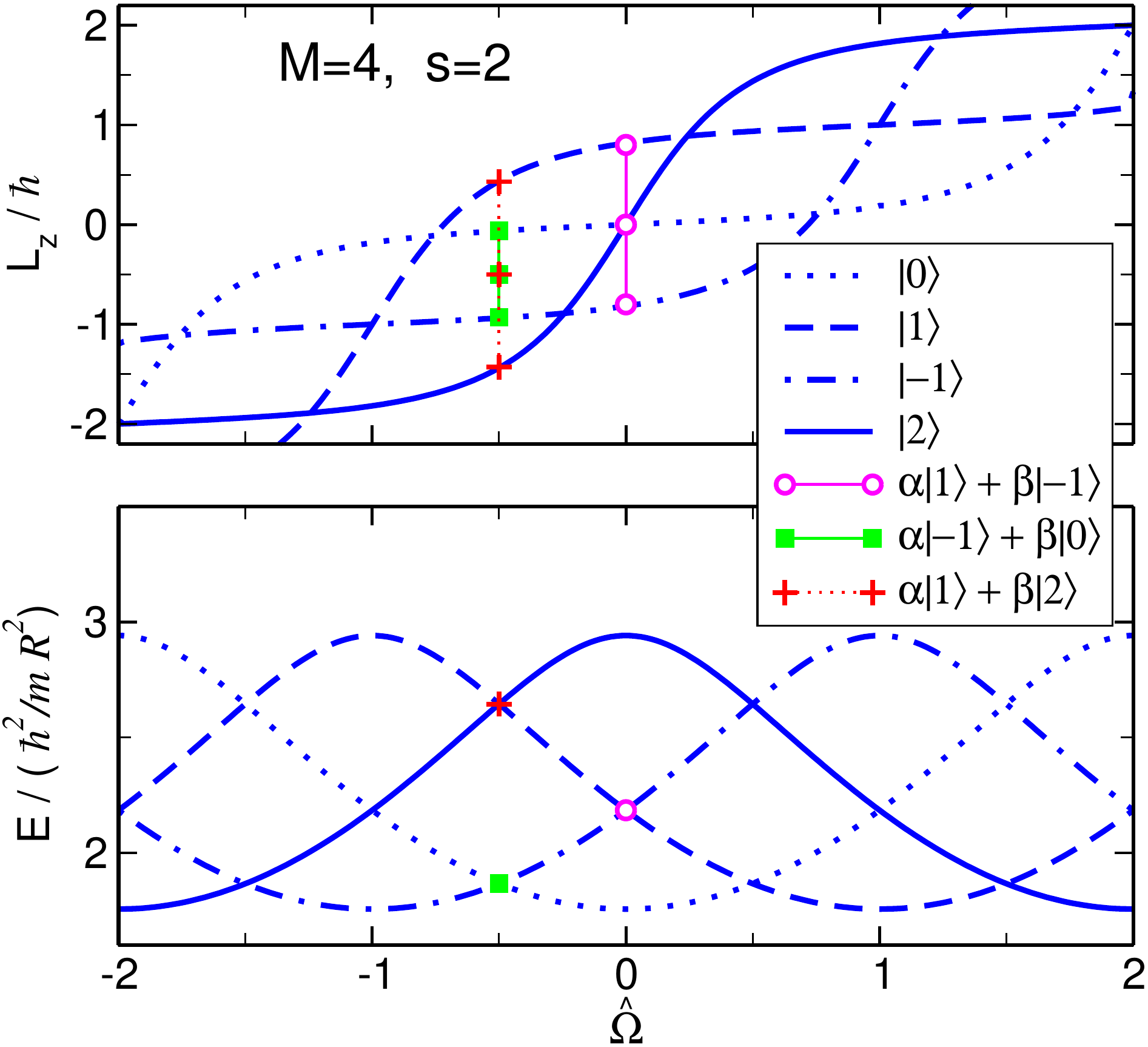}
 \caption{Energy (lower panel) and mean angular momentum (upper panel) of 
linear 
Bloch states in the first energy band of a 
ring lattice rotating at rate $\hat\Omega=mR^2\Omega/\hbar$ and 
same parameters as the right panel of Fig.\ \ref{fig:M34band}. The symbols, 
overlapped in the energy graph and connected by thin vertical lines 
in the angular momentum graph, indicate three particular cases of degenerate 
states.}
 \label{fig:linearM4}
\end{figure}
As shown in Fig.\ \ref{fig:M34band}, all the  
eigenstates in each energy band except those lying 
at the center $q=0$ and edge 
$q=M/2$ of the Brillouin zone, are doubly degenerate.
As a consequence of this degeneracy, new eigenstates of non definite 
quasimomentum can be obtained by linear superposition of 
definite-quasimomentum Bloch waves, that is 
$\psi=\alpha|n,q\rangle+\beta|n,-q\rangle$, where $\alpha$ and $\beta$ are 
complex coefficients satisfying $|\alpha|^2+|\beta|^2=1$. Figure 
\ref{fig:linearM4} highlights a representative case at $\hat\Omega=0$ inside 
the 
first energy band of a 4-site lattice (open symbols). The two panels collect 
data from the numerical solution of Eq.\ (\ref{eq:1DGP}) with $\hat g=0$. 
Whereas the degeneracy of states with quasimomenta $q=1$ and $q=-1$ is 
indicated by a single symbol in the energy graph (lower panel of Fig.\ 
\ref{fig:linearM4}), it corresponds to a continuum of states 
$\alpha|1\rangle+\beta|-1\rangle$ with 
mean angular momentum 
$L_z\sim\hbar(|\alpha|^2-|\beta|^2)$ in the interval 
$\hbar [-1,1]$ (represented by three connected symbols in the upper panel of 
Fig.\ \ref{fig:linearM4}).  As we will see, all the states in the 
mentioned interval find continuation in the nonlinear regime. The states with 
definite quasimomentum $q$ become the nonlinear Bloch waves, whereas those 
without definite $q$ transform into solitonic states.

The linear superpositions of degenerate Bloch waves give rise to new symmetries 
that do not share the usual periodic features of the Bloch states.
As an example, taken from the ring lattice with $M=3$ of Fig.\ 
\ref{fig:M34band}, the linear combinations 
of the degenerate eigenstates $\psi_{1,1}\equiv|1,1\rangle$ and 
$\psi_{1,-1}\equiv|1,-1\rangle$ give rise to states with one or two density 
peaks, for the symmetric $|1,1\rangle+|1,-1\rangle$ and the antisymmetric 
$|1,1\rangle-|1,-1\rangle$ cases respectively (see Fig.\ \ref{fig:M34wf} for the 
continuation of these states in the nonlinear regime). As we will show 
below, these two different symmetries, having zero mean angular 
momentum, are manifested in two separated families of nonlinear, soliton-like 
states. Furthermore, for other linear combinations of Bloch waves (other 
than the symmetric and the antisymmetric in this case) with nonzero mean 
angular momentum, it is not  possible to find an equivalent nonlinear 
continuation by keeping $\hat\Omega=0$. Such nonlinear continuation does still 
exist for nonzero rotation rate $\hat\Omega\neq0$.

\subsection{Rotation}

Additional energy degeneracies are induced by rotation. To see this, we first 
notice that the few-site rotating lattice maps to the infinite lattice due to the fact 
that the Gross-Pitaevskii equation for the Bloch wave amplitude 
${\hat u_{n,q}}(\theta)$ in the rotating frame
\begin{equation}
  \left\lbrace
\frac{1}{2}\left(-i\partial_\theta+q-\hat \Omega \right)^2
  + \hat V(\theta)+  
\hat{g} |{\hat u_{n,q}}|^2
  - \hat{\mu}\right\rbrace{\hat u_{n,q}} = 0\, .
  \label{eq:1DGPq}
\end{equation}
is also the equation of motion of a Bloch state with quasimomentum 
$q'=q-\hat\Omega$ in the static infinite lattice. 
In particular, in the noninteracting linear regime ($\hat{g}\rightarrow 0$) the energy
$E_q(\hat\Omega)$ (in units of $E_R$)
of a Bloch wave of (discrete) quasimomentum $q$ in the 
rotating ring lattice coincides with the energy $E^\infty(q-\hat\Omega)$ of a Bloch 
wave of quasimomentum $q'=q-\hat\Omega$ in the static infinite lattice. 
Therefore, for fixed $q$, by varying conveniently the parameter $\hat \Omega$ 
one can expand the eigenstates of the infinite lattice in the whole Brillouin zone 
$q'\in(-\pi/d,\pi/d]$. 
Since the group velocity in the infinite lattice is given by 
$\hat v=\partial_{q'} E^\infty(q')$ \cite{Kittel1996}, the above equivalence enables 
us to infer the superfluid velocity in the rotating ring lattice as a function of the
angular velocity
\begin{align}
\hat v(\hat\Omega)=-\partial_{\hat\Omega} E_q(\hat\Omega).
\label{eq:group_velo}
\end{align}

As $\hat\Omega$ varies, the Bloch states $|q\rangle$ (labeled by their 
quasimomentum $q$ at $\hat\Omega=0$) shift their positions in the energy graph 
(see lower panel of Fig.\ \ref{fig:linearM4}) according to the periodicity
introduced by both the ring, $\Delta \hat\Omega=M$, and the lattice 
$\Delta \hat\Omega =1$.
Then, in the Brillouin zone $E_q(\hat\Omega)=E_q(\hat\Omega + l\,M)$, with $l$ 
integer. Besides, consecutive Bloch waves $|q\rangle$ and $|q+1\rangle$ swap 
their energies after a $\Delta \hat\Omega=1$ interval in the angular rotation, 
such that $E_q(\hat\Omega=q+1)=E_{q+1}(\hat\Omega=q)$. In general, for each 
pair of Bloch waves with $(q_1,\,q_2)$, these symmetries imply energy 
degeneracies at $\hat\Omega= (q_1+q_2)/2$ and  $\hat\Omega= (q_1+q_2 + M)/2$, 
that is at 
either integer or half-integer $\hat\Omega$. The total number of 
degeneracies per energy band is the the number of combinations $\binom{M} 
{2}$. Two examples are explicitly indicated in Fig.\ \ref{fig:linearM4}. 
Symbols at $\hat\Omega=-0.5$ mark the superpositions 
$\alpha|-1\rangle+\beta|0\rangle$ and $\alpha|1\rangle+\beta|2\rangle$. Again, 
the degenerate states expand continuum intervals in the graph of the mean 
angular momentum per particle (upper panel of Fig.\ \ref{fig:linearM4}),
with $L_z\sim\hbar[-1,0]$ and $L_z\sim\hbar[-1.5,0.5]$ respectively. 

Taking into account Eq.\ (\ref{eq:current}), the quantization of the circulation 
around the ring lattice in units of the ``flux quantum'' $\phi_0=2\pi\hbar/m$ 
takes the form
\begin{align}
\hat\Gamma=\frac{1}{2\pi}\oint d\theta \,\hat p(\theta)=\frac{1}{2\pi}\oint 
d\theta \,\left(\hat v + \hat \Omega\right) = l \,,
\label{eq:circulation}
\end{align}
with $l=0,\,1,\,2\dots$. This expression is clearly equivalent to the well 
known fluxoid quantization in superconductivity \cite{Barone1982}. 
It amounts to encircle $l$ vortices inside the ring, for the phase jumps in 
$2\pi\,l$ when encircling the system. Singular situations arise if a vortex 
lays exactly on the ring, producing a node on the corresponding wave 
function. This occurs at a particular value of the rotation rate in the transit 
of a vortex from the outer to the inner regions of the ring or vice versa.
Since each vortex (i.e.\ each node) on the ring imposes a sudden $\pi$-phase 
jump at its location, the phase winding around the ring now leads to
\begin{align}
\frac{1}{2\pi}\, P\! \oint d\theta \,\hat p(\theta)=\frac{1}{2\pi}\, P\! \oint 
d\theta \,\left(\hat v + \hat \Omega\right) = \frac{\nu}{2}\,,
 \label{eq:half_Lz}
\end{align}
where $\nu$ is the number of nodes and $P$ denotes the Cauchy principal 
value. Thus the presence of nodes in the wave function extends the 
quantization of circulation Eq.\ (\ref{eq:circulation}) to half integers of $h/m$.

From Eq.\ (\ref{eq:current}), when the superfluid velocity vanishes, then 
$\hat p(\theta)\equiv\partial_\theta \arg(\psi)=\hat \Omega$ and hence, 
according to the quantization of circulation, the phase has to be linear,
$\arg(\psi)=j\, \theta$, with $j$ being an integer or half-integer.
For a Bloch wave $|q\rangle$, from Eq.\ (\ref{eq:group_velo}), the superfluid 
velocity vanishes at the minima and maxima of the dispersion relation 
$E_q(\hat\Omega)$ in the rotating frame. The minimum corresponds to the 
ground state, at $\hat\Omega=q$ (see, for example, the lower panel of 
Fig.\ \ref{fig:linearM4}), it has no nodes and the linear phase takes the value 
$\arg(\psi)=q\, \theta$. Substituting these results in Eq.\ (\ref{eq:Lz}) and 
making use of the normalization condition one finds that at the minimum
of the dispersion relation $E_q(\hat\Omega)$ of a Bloch wave $|q\rangle$,
which occurs at $\hat\Omega=q$, the mean angular momentum per particle
takes the value $L_z=\hbar\,q$, corresponding to the presence of $q$
vortices inside the ring lattice.

For a Bloch wave $|q\rangle$, the superfluid velocity also vanishes at the energy 
maxima, on the edge of the Brillouin zone $\hat\Omega=q+M/2$. In this case 
the phase reads $\arg(\psi)=(q+M/2)\, \theta$, indicating the presence of $M$ 
nodes on the ring. The mean angular momentum per particle takes correspondingly 
half integer (for M odd) or integer (for M even) values of $\hbar$ (see e.g.\ 
Fig.\ \ref{fig:linearM4}). 
In the absence of rotation, the single-value property of the wave 
functions precludes the fulfillment of Eq.\ (\ref{eq:half_Lz}) for odd number of 
nodes.

The above results lead to the following picture for a rotating ring lattice
with $M$ sites (see e.g.\ Fig.\ \ref{fig:linearM4}). Starting from the ground
state of a Bloch wave $|q\rangle$ in the first 
energy band (which occurs at $\hat\Omega=q$ and has a mean angular 
momentum per particle $L_z=\hbar\,q$) as $\hat\Omega$ increases adiabatically 
so does the condensate energy $E_q(\hat\Omega)$ while $L_z$ remains almost 
unaltered reflecting the superfluidity of the system. This situation holds 
until $E_q(\hat\Omega)$ approaches a maximum, which occurs at 
$\hat\Omega=q+M/2$ and corresponds to the presence of $M$ vortices lying 
exactly on the ring lattice. At this point $L_z$, which increases suddenly in 
the vicinity of this configuration, takes the value $L_z=q+M/2$. Finally, as 
$\hat\Omega$ increases further the above $M$ vortices enter the lattice 
and $L_z$ keeps increasing up to $L_z=q+M$. This picture essentially remains 
true in the nonlinear regime.

\begin{figure}[t]
 \centering
 \includegraphics[width=\linewidth]{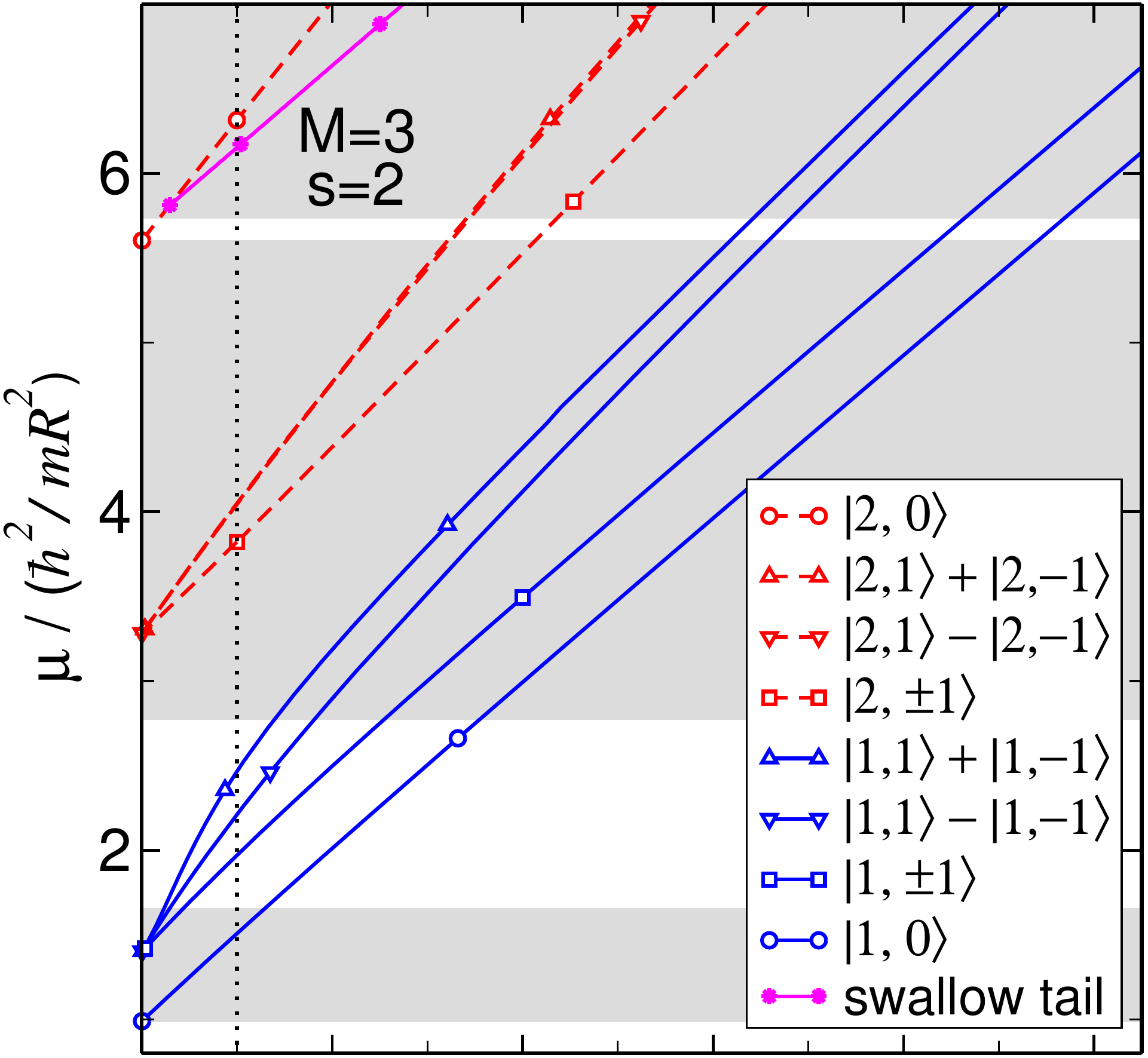} \\ \vspace{2mm}
 \includegraphics[width=\linewidth]{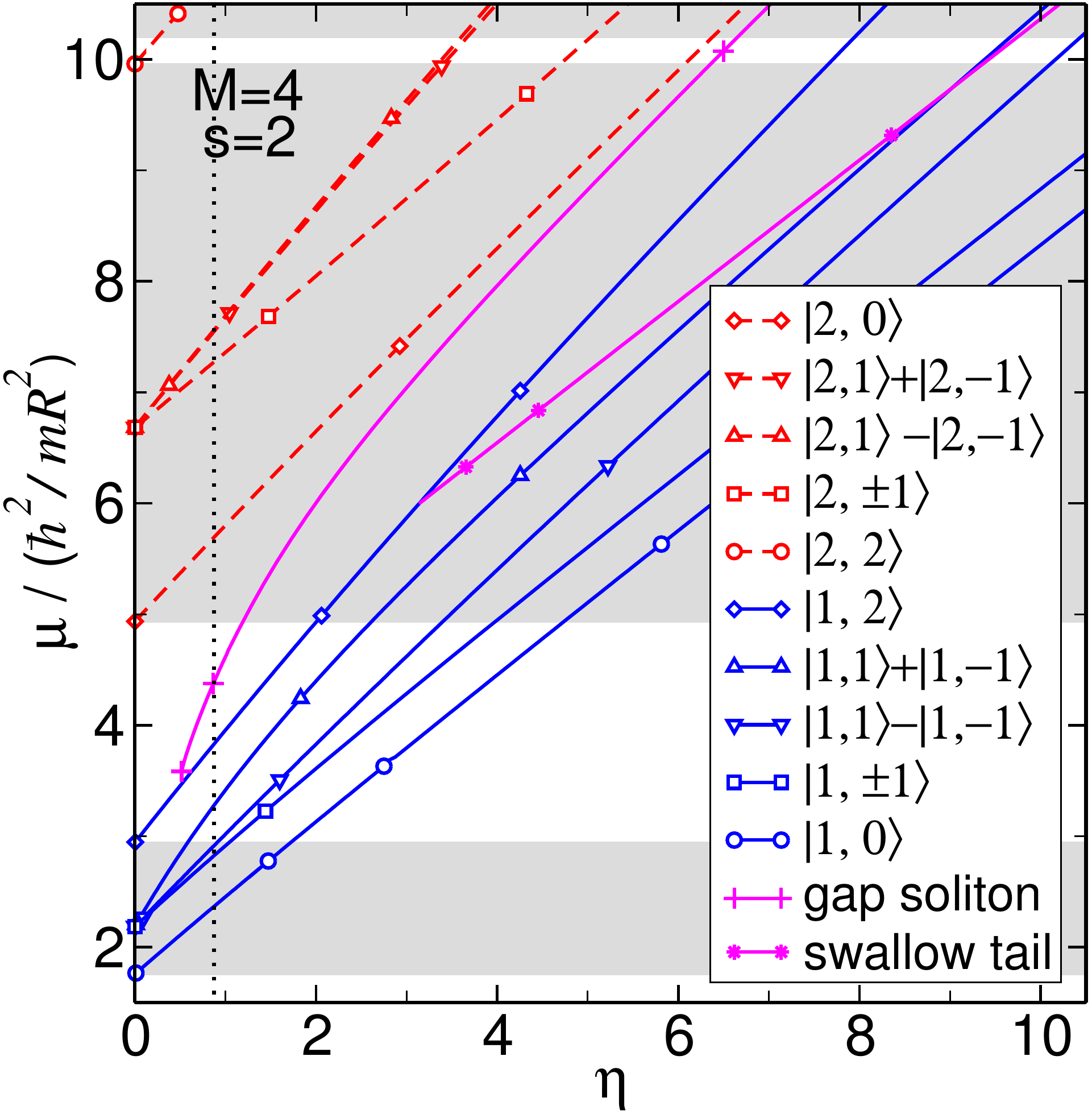}
 \caption{Trajectories in the chemical potential versus interaction ($\eta=\hat 
g N/M$) graph  of nonlinear states living in the ring lattices of Fig.\ 
\ref{fig:M34band} at $\hat\Omega=0$. The grey-shaded regions represent 
the linear energy bands of the infinite lattices. The vertical dotted lines 
indicate particular values of the nonlinearity ($\eta\sim 1$) for which the 
state density profiles are shown in Fig.\ \ref{fig:M34wf}.
}
 \label{fig:M34nonlinear}
\end{figure}

\section{Nonlinear waves}

A simple picture, based on the continuation of states from 
the non-interacting regime, allows for the generation of the different 
families of nonlinear waves in the ring lattice. Figure \ref{fig:M34nonlinear} 
shows the outcome of this approach for the lattices considered in 
Fig.\ \ref{fig:M34band}. Families of states proceeding from the first (solid 
lines) and second (dashed lines) energy bands are shown at $\hat\Omega=0$. Gap 
solitons have their origin at the mentioned energy degeneracies 
of the linear regime. This is the case for the families originating 
from the combinations $|1,1\rangle\pm |1,-1\rangle$ for $M=3$ (top 
panel of Fig.\ \ref{fig:M34nonlinear}). For $M=4$ (lower panel) 
the linear origin of the gap-soliton family is not apparent. It bifurcates 
from the top of the first energy band in the nonlinear regime. However, it 
can be tracked back up to the linear regime at $\hat\Omega=0.5$, where it 
originates from the combinations $|1,2\rangle\pm |1,1\rangle$.

In general, an increasing interaction strength leads the nonlinear 
states to resonate with excitation modes associated with the underlying linear 
energy spectrum. These resonances are responsible for the generation of 
pitchfork bifurcations that translate into swallow tails at the maxima of the 
nonlinear energy bands (shown in Fig.\ \ref{fig:M34nonlinear}, at the second 
band for $M=3$ and at the first band for $M=4$ ), or saddle node bifurcations 
of gap solitons in higher energy gaps. In what follows, we provide evidence for 
these statements.

\begin{figure}[tb]
 \centering
 \includegraphics[width=\linewidth]{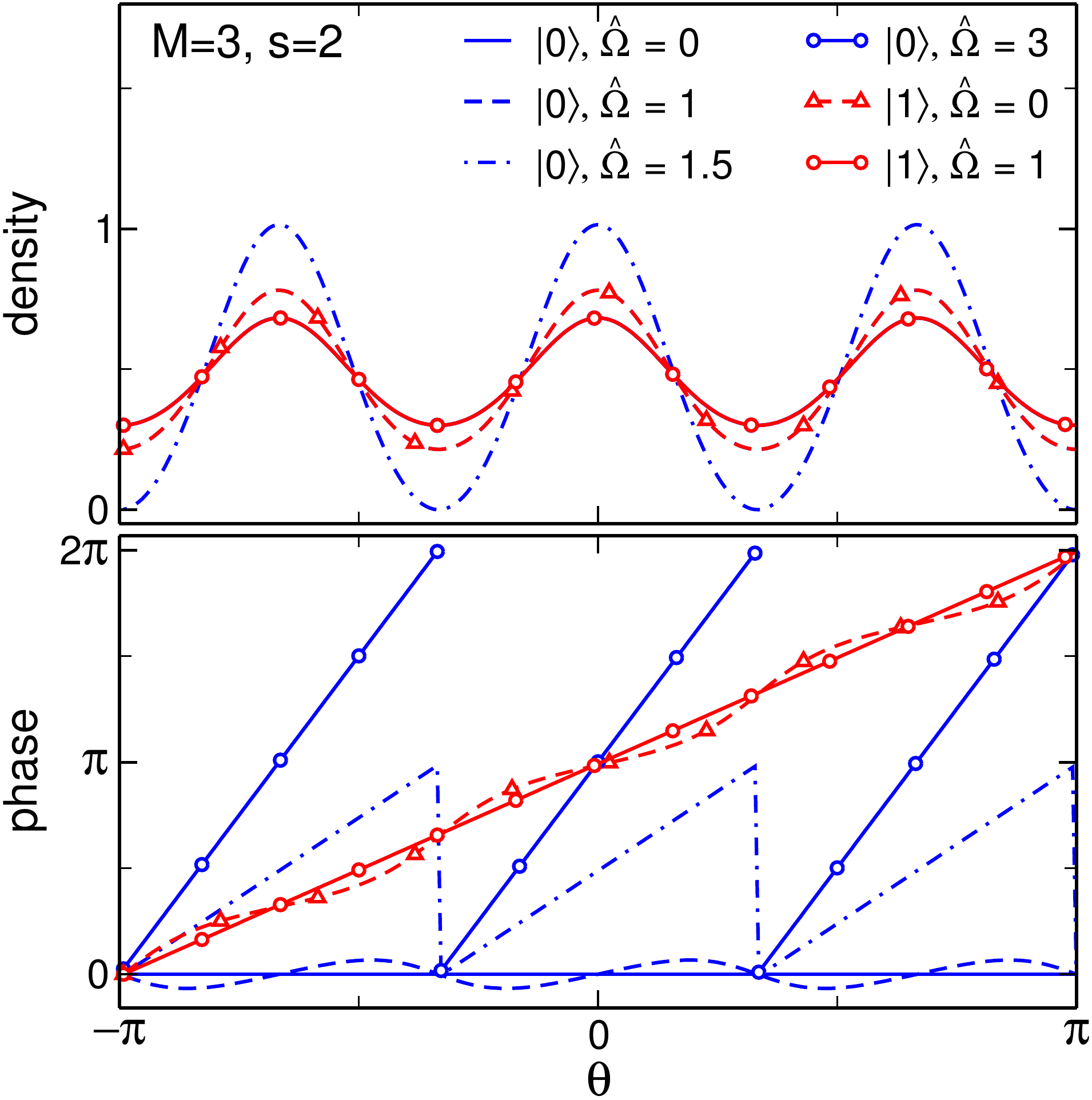}
 \caption{Number density and phase profiles of the nonlinear Bloch 
waves $|0\rangle \equiv |1,0\rangle$ and $|1\rangle \equiv 
|1,1\rangle$ in  the $M=3$ ring lattice of Fig.\ \ref{fig:M34band}
 for different rotation values $\hat\Omega$.
The phase winding induced by rotation causes  $|0\rangle$ to evolve 
from a non-current state at 
$\hat\Omega=0$, through a soliton-train-like state at $\hat\Omega=1.5$, into a
vortex-like (ring-current) state at $\hat\Omega=3$.}
 \label{fig:M3phase}
\end{figure}
\begin{figure}[t]
 \centering
 \includegraphics[width=\linewidth]{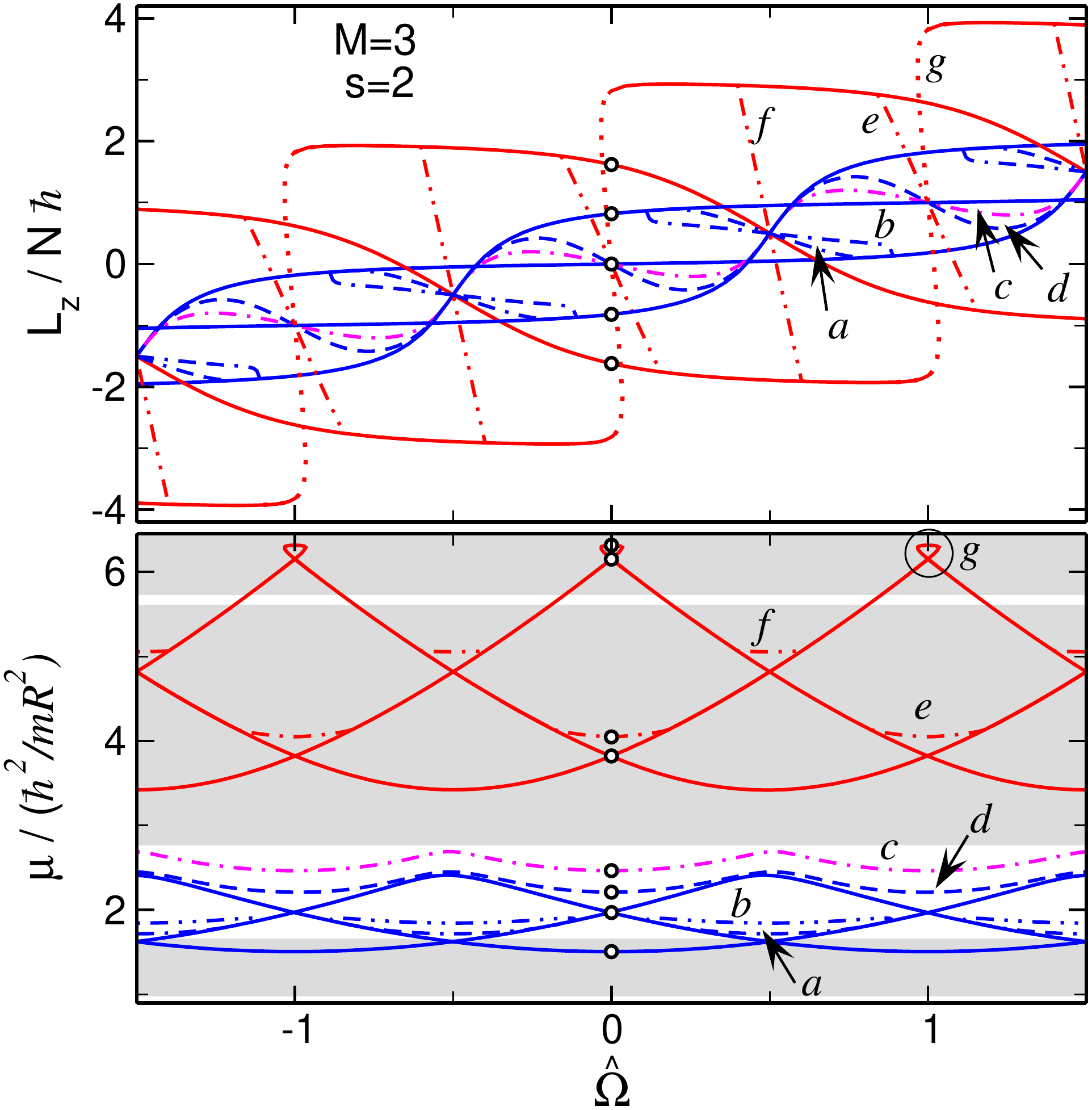}
 \caption{Mean angular momentum per 
particle (top panel) and chemical potential (bottom panel) of stationary states
in the first and second energy bands of a rotating ring lattice with 
$M=3$, $s=2$, and $\eta=1$. Open symbols indicate the non-rotating eigenstates 
depicted in the left panels of Fig.\ \ref{fig:M34wf}. The shaded regions 
correspond to the linear energy bands, and the solid lines represent the 
Bloch states. The discontinuous lines correspond to nonlinear 
states without definite quasimomentum. For instance, states of the families 
labeled by $a$ and $b$, in the first band, and $e$ and $f$, in the second band, 
form intra-band swallowtails. The families $c$ and $d$ are made of 
gap-soliton-like states, whereas the (encircled) states $g$ form an out-of-band 
swallowtail.
}
 \label{fig:M34omega}
\end{figure}
\subsection{Persistent current switch}

The nonlinear families of Bloch waves support supercurrents that 
can be controlled by means of rotation. In Fig.\ \ref{fig:M3phase}, we have 
summarized the changing configuration (density and phase) of two Bloch waves, 
$|0\rangle \equiv |1,0\rangle$ and $|1\rangle \equiv |1,1\rangle$,
with varying angular rotation. The system parameters are the same as in the 
3-site lattice of Fig.\ \ref{fig:M34band}, but now with an interaction parameter 
$\eta=1$. The energy and mean angular momentum of these Bloch waves can be also 
compared with other stationary states in Fig.\ \ref{fig:M34omega}. Monitoring 
the phase of state $|0\rangle$ in the transit from $\hat\Omega=0$ to 
$\hat\Omega=3$ (lower panel Fig.\ \ref{fig:M3phase}), one 
can see that three nodes make simultaneously their appearance at 
$\hat\Omega=1.5$ (dot-dashed lines), just when the state reaches its maximum
chemical potential as a function of $\hat\Omega$ (see 
Fig.\ \ref{fig:M34omega}). At this point the circulation due to the presence of 
three nodes is $\hat\Gamma=3/2$. In the transit from 
$\hat \Omega <1.5$ to $\hat \Omega >1.5$ the circulation jumps from 
$\hat\Gamma=0$ to $\hat\Gamma=3$, indicating the entry of three 
vortices inside the ring. The mean angular momentum keeps increasing with 
$\hat\Omega$ from $L_z=0$ at $\hat\Omega=0$ up to $L_z=3\hbar$ at 
$\hat\Omega=3$, where the density profile matches the original configuration at 
$\hat\Omega=0$.

The Bloch state $|1\rangle$ makes an equivalent 
transit from  $\hat\Omega=1$ to $\hat\Omega=4$. The density profiles are 
exactly the same as those of $|0\rangle$ at one less unit of $\hat\Omega$. The 
phase profiles of state $|1\rangle$, instead, reflect one additional unit 
of circulation. The mean angular momentum increases from $L_z=\hbar$ to 
$L_z=4\hbar$.

This smooth variation of Bloch waves suggests an experimental 
procedure to switch between persistent currents in toroidal 
systems by means of ring lattices. The advantage with respect to previous 
methods using Gaussian weak links to drag a BEC \cite{Eckel2014} is the 
absence of sudden, uncontrolled phase slips. By using an 
$M$-site lattice, persistent currents that differ in $M$ winding numbers can be 
smoothly switched. Nevertheless, it is important to add 
that the Bogoliubov analysis of the states making such connections 
reveals dynamical instabilities for intermediate values of $\hat\Omega$. For 
instance, in the particular case shown in Fig.\ \ref{fig:M3phase}, and within 
the transit of $|0\rangle$ in the range $\hat\Omega\in[0,3]$, the intermediate 
states in the window $\hat\Omega\in[0.92,2.08]$ are unstable. However, a smooth
transit, faster than the typical growth rate of the unstable modes, 
through these intermediate regions could still adiabatically (passing through 
all the 
intermediate states of the Bloch wave) lead the system into the final
stable current states. This transition is not qualitatively different to the 
driven transition of a linear system in the observation of Bloch 
oscillations.

\begin{figure}[tb]
 \centering
 \includegraphics[width=0.515\linewidth]{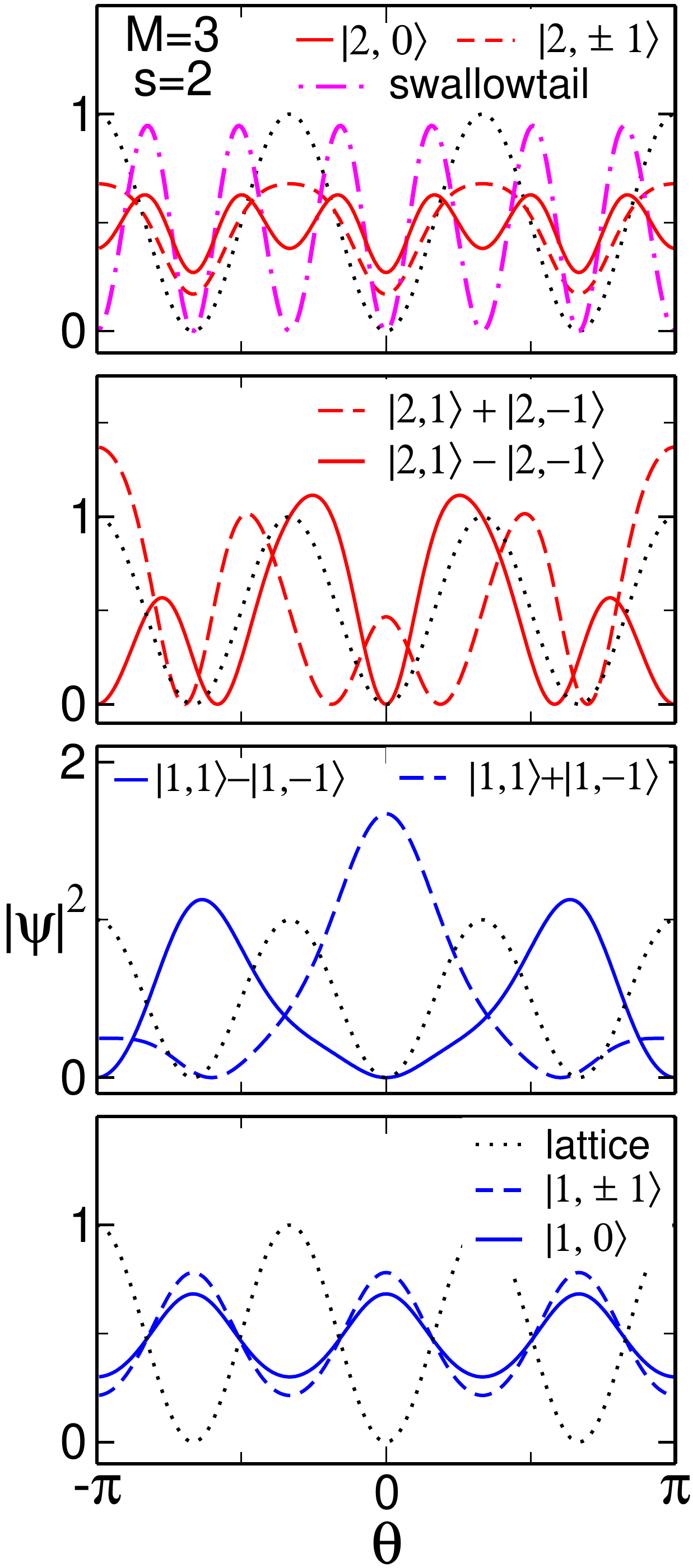} 
 \includegraphics[width=0.472\linewidth]{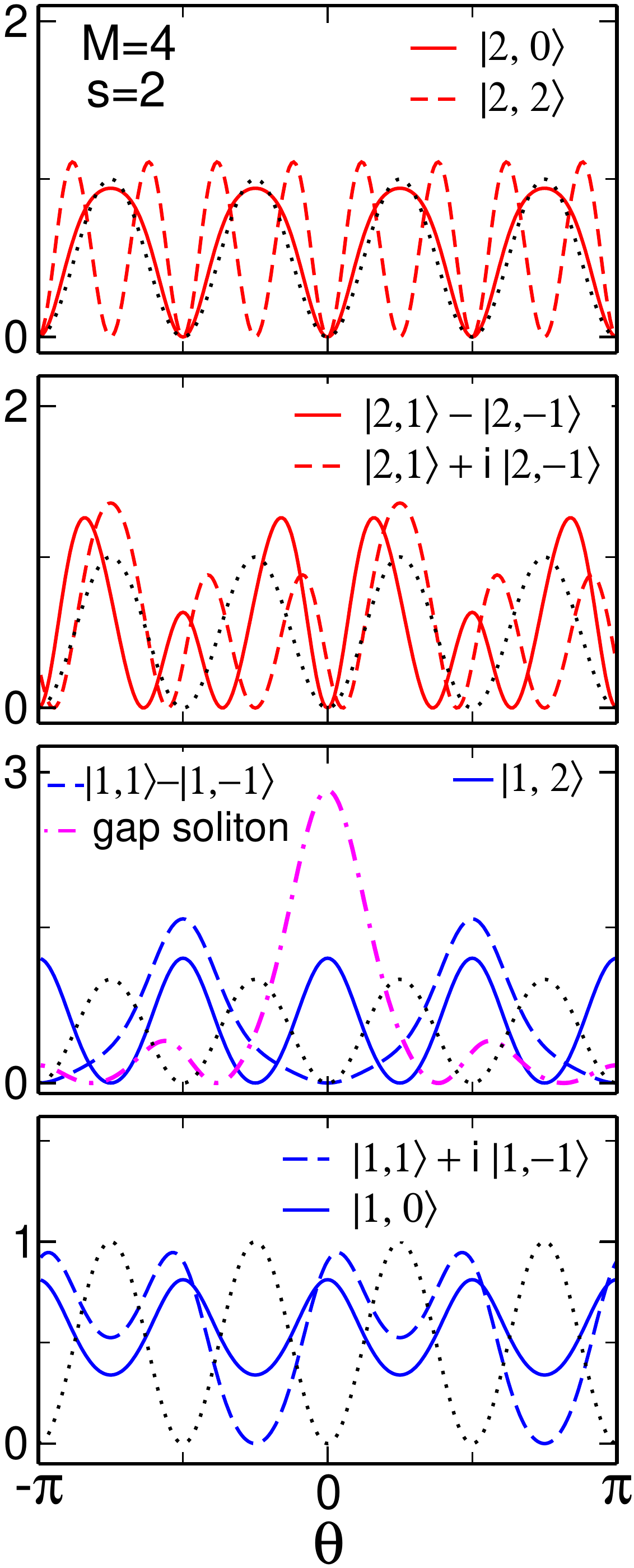}
 \caption{Azimuthal densities of the stationary 
states of the GP Eq.\ (\ref{eq:1DGP}) with $\hat\Omega=0$ 
in  the ring lattices of Fig.\ \ref{fig:M34band} and 
nonlinearity given by the vertical dotted lines in Fig.\ 
\ref{fig:M34nonlinear}. The dotted lines represent the lattice profile. 
}
\label{fig:M34wf}
\end{figure}

\subsection{Dark-soliton-like states}

\begin{figure}[t]
 \centering
 \includegraphics[width=\linewidth]{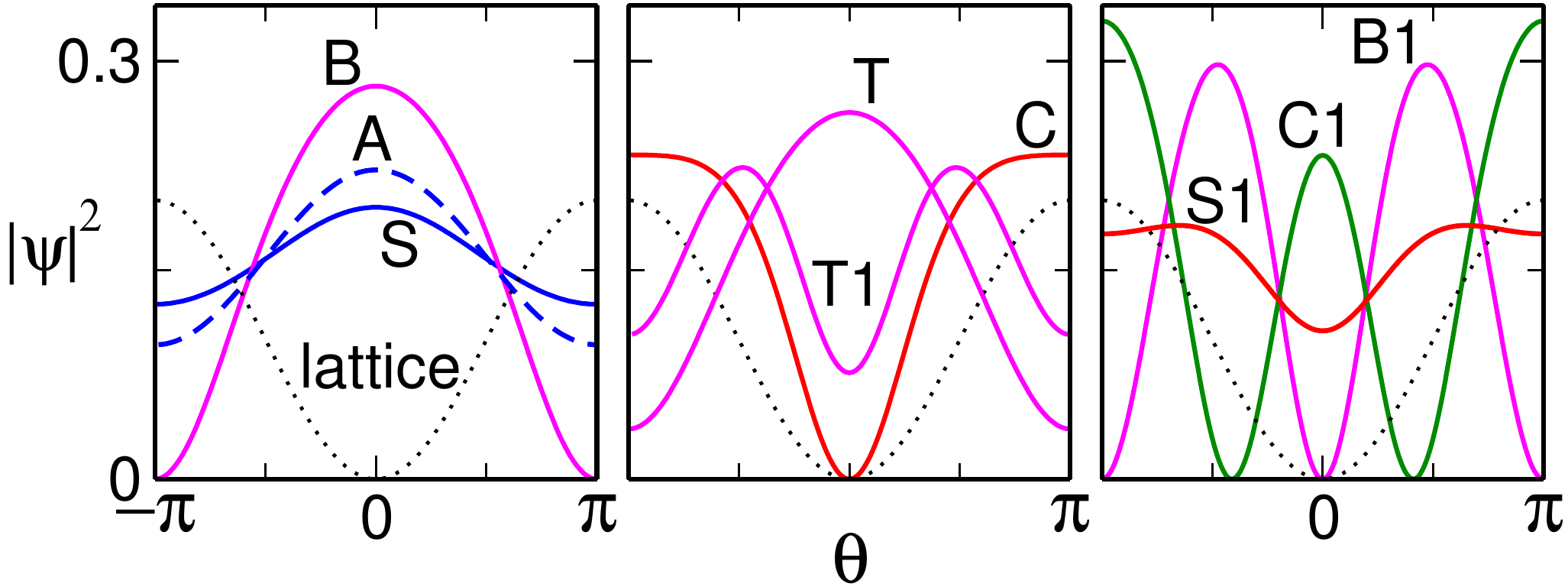}\\ \vspace{3mm}
 \includegraphics[width=\linewidth]{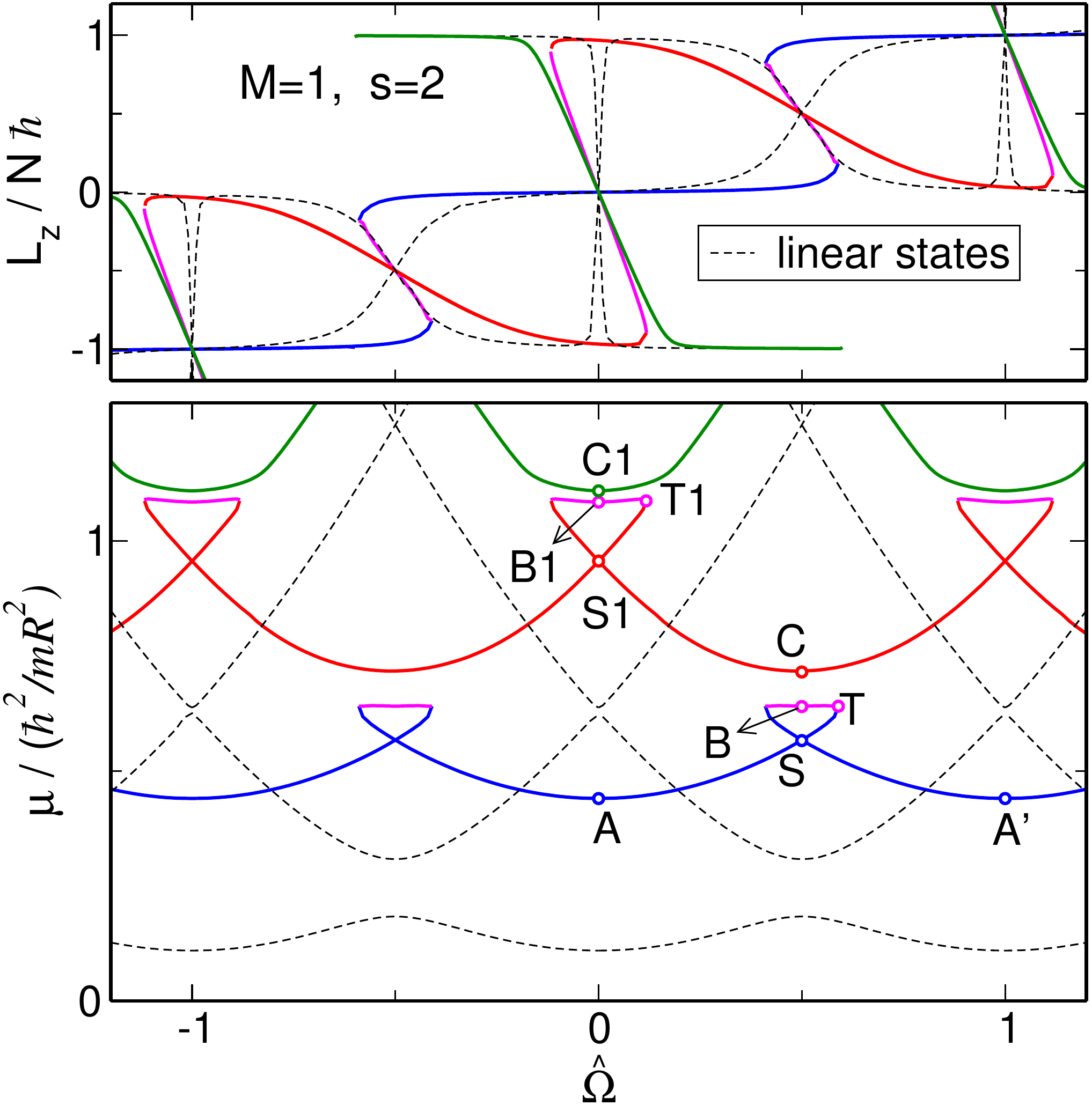}  
 \caption{Stationary states of the Gross-Pitaevskii Eq.\ (\ref{eq:1DGP}) with 
interaction parameter $\eta=2$ for the first energy bands of a rotating ring 
lattice with $M=1$ and $s=V_0/E_L=2$. Top panels: density profiles of labeled 
states. Lower panels: chemical potential (bottom) and mean angular momentum per 
particle (middle) in the first (blue lines), second (red lines) and third 
(green lines) energy bands. Families of soliton-like states (magenta lines) 
form the out-of-band swallow tails in the first and second energy bands.
}
 \label{fig:M1omega}
\end{figure}

The most striking difference in the energy bands with respect to the linear 
regime is the generation of swallowtails beyond a threshold 
value of the interaction strength \cite{Wu2000,Diakonov2002,Machholm2003}. 
These structures of the energy bands break 
the smooth connection described before between stable states supporting 
persistent currents. The swallowtails are shaped at the 
local maxima of the nonlinear energy bands, as can be seen (encircled) in 
the second band of the bottom panel of Fig.\ \ref{fig:M34omega}, and more
clearly in both the first and second energy bands in Fig. \ref{fig:M1omega}.
These structures, that will be referred as out-of-band swallowtails, are 
developed at the edges of the Brillouin zone in the lowest 
energy band when $g\bar n>sE_L/2$, where $\bar n=N/2\pi\,R$ is the 
average density, or at the center of the Brillouin zone in the first excited 
(second) energy band when $g\bar n> 4 E_L(\sqrt{1+(s/8)^2}-1)$ (as it is the 
case in the bottom panel of Fig.\ \ref{fig:M34omega}). In the ring units, these 
thresholds are $\eta_1=\pi M\, s/8$ and $\eta_2=\pi M\,(\sqrt{1+(s/8)^2}-1)$, 
respectively, with $\eta_2<\eta_1$.

Along with the out-of-band swallow tails, there are also swallow tails in the 
energy-band interior, as those shaped by the curves $a$, $b$, $e$ and 
$f$ in  Fig.\ \ref{fig:M34omega}. These structures, that we will refer as 
intra-band swallowtails, are made of periodic states that have different 
period $\lambda d$ ($\lambda$ integer) from the lattice \cite{Machholm2004}. 
They have been interpreted as states made of dark solitons, due to the presence 
of nodes in their wave functions (as can be seen in Fig.\ \ref{fig:M34wf}). As 
we demonstrate below, these families of nonlinear waves proceed from the 
combinations of linear degenerate states.

\subsubsection{Out-of-band swallowtails and sudden phase slips}
Let us first investigate the ring rotating ``lattice'' with just one site 
$M=1$. For weak periodic potentials, this minimal system can 
be seen as a toroidal condensate in the presence of a wide weak link. As a 
result the associated physics of persistent currents in a ring can be recovered 
\cite{Mateo2015}. Specifically, the case of a weak sinusoidal potential in a 
ring has been addressed from the point of view of the generation and stability 
of dark soliton states \cite{Fialko2012}.

In Fig.\ \ref{fig:M1omega} we show the chemical potential and the
mean angular velocity  of the stationary states with $s=2$ and interaction 
$\eta=2$. Since in this case $\eta_1=0.79$ and  $\eta_2=0.1$, there are swallow 
tails at the first and the second energy bands. As a consequence, the
ground state $|1,0\rangle$ at $\hat\Omega=0$,
indicated by the label A in the bottom panel of Fig.\ 
\ref{fig:M1omega}, cannot smoothly transit from $\hat\Omega=0$, where it has 
$L_z=0$, up to $\hat\Omega=1$, where it has $L_z=\hbar$ (point A'). The energy 
curve of state $|1,0\rangle$ is broken into equal pieces with different 
circulation centered around the integer values of $\hat \Omega$. Each piece 
represents a metastable supercurrent state that effectively stretches beyond 
the edge of the Brillouin zone, at half-integer values of $\hat \Omega$, where 
it is not the ground state. Eventually the energy path reaches a maximum at a 
critical rotation rate $\hat{\Omega}_c$ (point T), beyond which it does not 
exist. 
As a result, the adiabatic evolution through the swallow tail is not 
possible, and just changing $\Omega$ beyond the outer edge of the tail 
structure results in non-adiabatic dynamics, a sudden phase slip, which will 
also change the total circulation by $2\pi$. The path from A to T in 
Fig.\ \ref{fig:M1omega} can be done adiatically, but increasing $\Omega$ 
any further will cause the system to drop to the line connected to A' (plus 
phonon-like excitations). A similar dynamics takes place in the absence of 
lattice, as has been demonstrated in toroidal systems with a weak link 
\cite{Eckel2014,Mateo2015}, where an increase of rotation  beyond the 
swallowtail $\hat\Omega>\hat{\Omega}_c$ brings a metastable state through a 
sudden phase slip event, which changes the circulation, into the corresponding 
ground state.

The crossing energy paths of two (consecutive) metastable supercurrents are 
connected, at their maxima, by a family of solitonic states. The state B at 
$\hat\Omega=0.5$ (see Fig.\ \ref{fig:M1omega}) represents a clear example of 
the latter. The energy curves of these families trace the 
out-of-band swallow tails, that reflect the 
presence of hysteresis in the system \cite{Mueller2002,Eckel2014}.
\begin{figure}[tb]
 \centering
 \includegraphics[width=\linewidth]{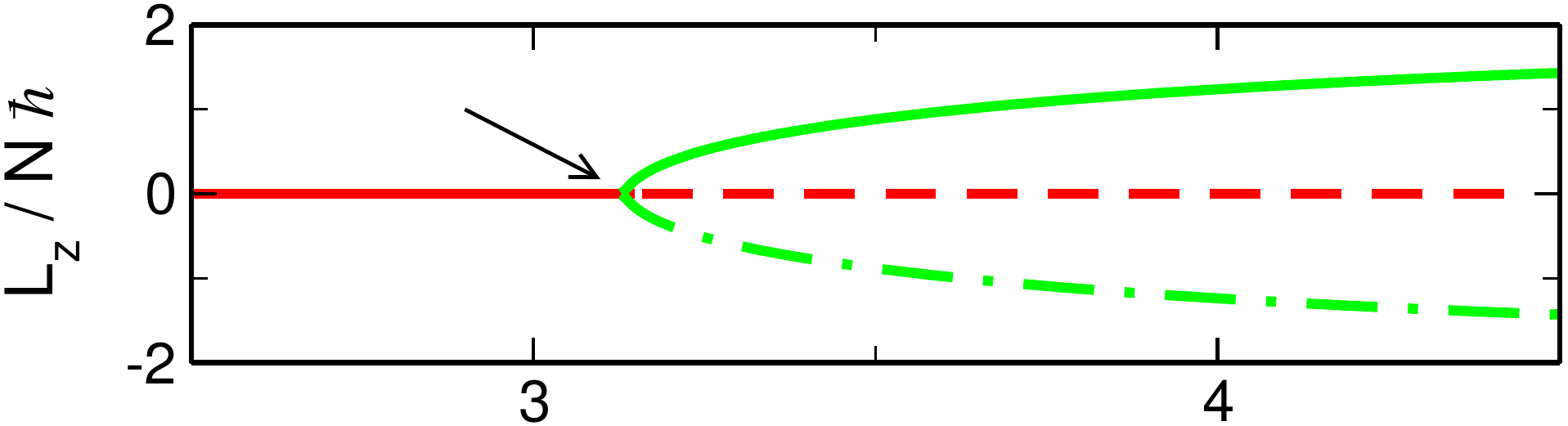}\\
 \includegraphics[width=\linewidth]{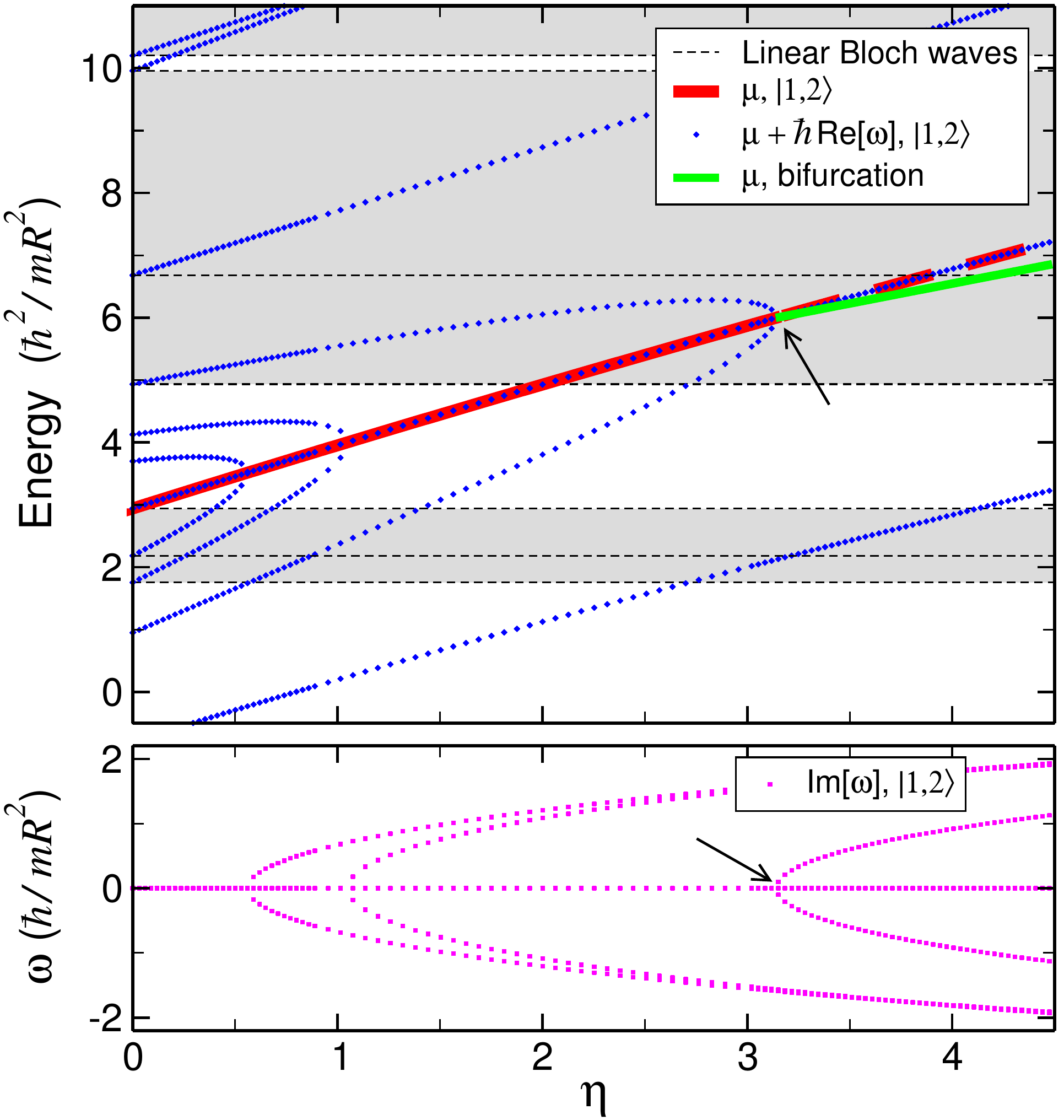}
 \caption{  Top panel: pitchfork bifurcation (marked by the arrow) 
of states $|1,2\rangle$ (thick red lines) generating an out-of-band 
swallowtail in the first energy band of a ring lattice with $M=4$, $s=2$, and 
$\hat\Omega=0$. Two new current states with the same 
chemical potential (thick green curves) emerge at interaction $\eta_1=\pi$.
Middle panel: Chemical potential of the solutions (thick lines) and, 
superimposed, sum of chemical potential and real part of Bogoliubov 
excitation energies (symbols) as a function of the interaction. Bottom panel:
Imaginary part of the frequencies of Bogoliubov excitations as a function of 
interaction.
}
 \label{fig:M4pitchfork}
\end{figure}
It is interesting to compare the states living on the swallow tails, e.g. 
points B and B1 in Fig.\ \ref{fig:M1omega}, with the corresponding states on 
the 
next energy band for the same rotation rate $\hat\Omega$, i.e. states C 
and C1, respectively. They present the same number of nodes. 
While state B has one node at the maximum of the periodic potential, state C 
has one node at the minimum. Similarly in the second band, B1 presents two 
nodes, one of which lays at the potential maximum, whereas C1 has both nodes 
within the potential well. As it was noted in Ref. \cite{Fialko2012}, this 
difference is crucial for the dynamical stability of the states, in 
such a way that those states having nodes at the maxima are 
unstable, but those with nodes inside the potential well are dynamically 
stable.

The out-of-band swallow tails are the result of a pitchfork bifurcation of the 
family of
nonlinear Bloch waves situated at the upper edge of the energy bands. To 
illustrate this point, we make use of a system with $M=4$, $s=2$, 
$\hat\Omega=0$, 
and varying interaction $\eta$, where swallow tails emerge in the first energy 
band at $\eta_1=\pi$. Figure \ref{fig:M4pitchfork} shows the excitation modes 
(dots) of the familiy of states $|1,2\rangle$ (thick red lines), which 
originates at the top of the first (linear) energy band. The excitation modes
have also continuation from the linear regime, where they coincide with the 
eigenstates of the linear Hamiltonian. As maxima of the energy band, the states 
$|1,2\rangle$ have zero angular momentum (see the top panel of Fig.\ 
\ref{fig:M4pitchfork}). However, at $\eta=\eta_1$ (indicated by arrows in the 
graphs) two new states with opposite currents bifurcate due to the resonance 
with an excitation mode originating from the second 
energy band (see the middle panel of Fig.\ \ref{fig:M4pitchfork}).

For higher interaction strength $\eta>\eta_1$ the family $|1,2\rangle$ includes 
new excitation modes with purely imaginary frequencies (bottom panel of Fig.\ 
\ref{fig:M4pitchfork}), denoting its dynamical instability. The bifurcating 
current-carrying states inherit the stability properties of the parent state 
before the bifurcation. In the present case they are also dynamically unstable 
due to resonances that take place at lower interaction strength 
$\eta<\eta_1$  (see lower panels of Fig.\ \ref{fig:M4pitchfork}) with 
excitation modes proceeding from the first linear energy band.

\subsubsection{Intra-band swallowtails}

For $M>1$, additional swallow tails originate from the connection between 
two crossing paths (in the $\mu$--$\hat \Omega$ graph) of nonlinear Bloch waves 
having different quasimomentum. The system shown in
Fig.\ \ref{fig:M34omega}, 
a ring lattice with $M=3$, $s=2$, and $\eta=1$, shows multiple examples of 
these structures along with small out-of-band swallow tails in the second 
energy band. 
Contrary to the latter, the intra-band swallowtails exist for 
arbitrarily small values of the interaction and originate from the energy 
degeneracies in the non-interacting regime. They clearly resemble the solitonic 
curves between winding number states found in the absence of lattice 
\cite{Kanamoto2009}.  But unlike such case, there are two families of solitonic 
states (and not only one, although hard to distinguish at the scale of 
Fig.\ \ref{fig:M34omega}) making the connection between states with definite 
quasimomentum on the upper part of each crossing. The number of these (double) 
swallow tails, per energy band, equals the number of energy degeneracies (or 
crossings) between pairs of the $M$ Bloch waves. 

This scenario is a generalization of the particular cases that have been 
presented in the literature \cite{Machholm2004}. In order to make the 
connection between the two states of different quasimomentum,  a number of 
nodes matching the difference in the circulation $\Delta \hat \Gamma $ are 
generated on the 
ring. The nodes appear in the states situated just at the middle of the swallow 
tails, and evolve into density depletions that become shallower as the 
solitonic states approach the merging point with the Bloch states.
The presence of the nodes breaks the symmetric pattern of the 
lattice and give rise to new periods.

A remark about the solitonic character of the states living on the 
described swallow tails is in order.  Although a general classification of 
these states as trains of dark solitons can be done, as regarding the 
presence of nodes, they can also be considered trains of bright solitons, 
as regarding the localization of the particle density. The next subsection 
clarifies this point.

\subsection{Bright solitons}

\begin{figure}[tb]
 \centering
 \includegraphics[width=\linewidth]{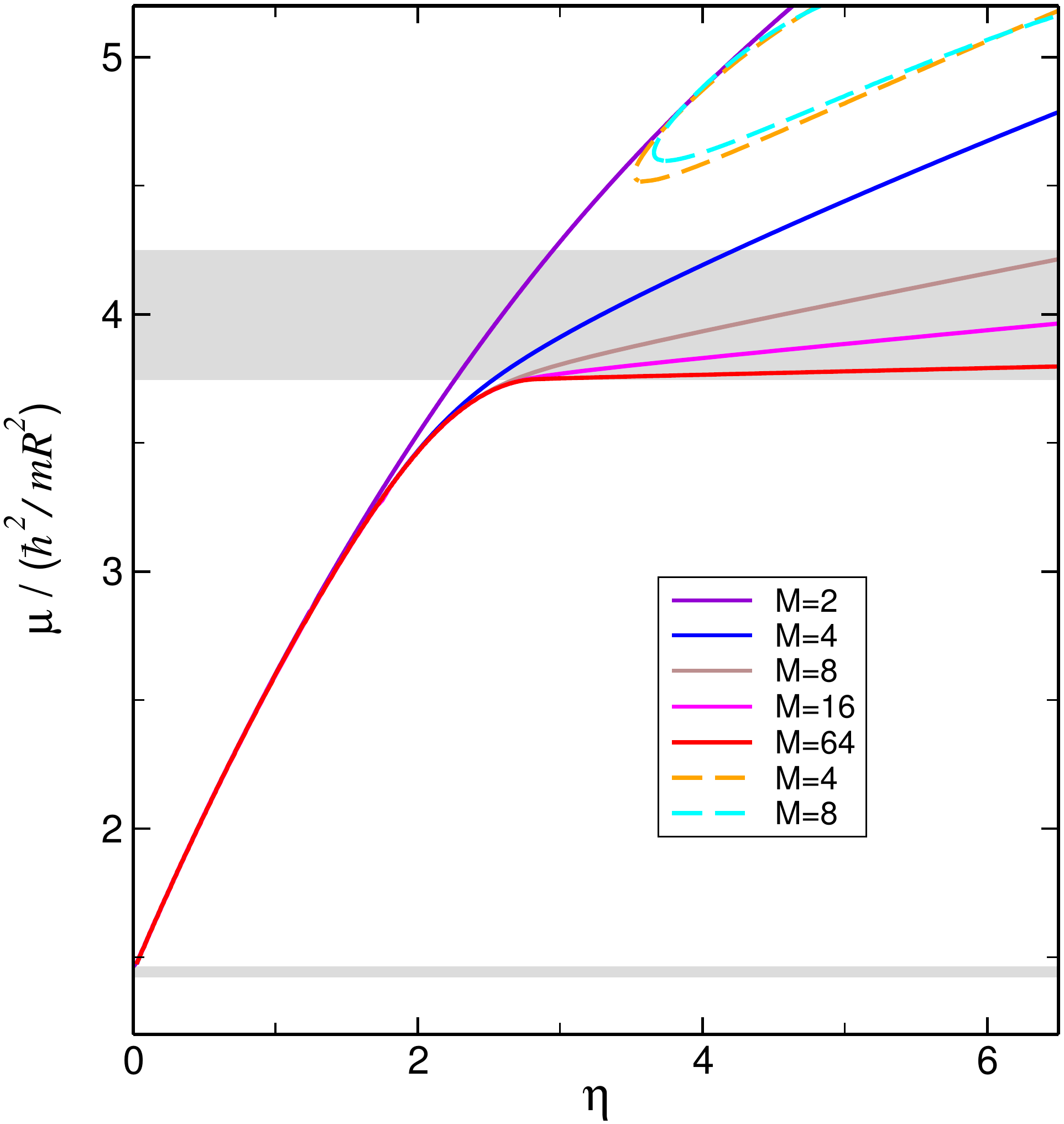}
 \caption{Chemical potential versus interaction strength of gap solitons in 
ring 
lattices with varying number of sites $M$ but the same depth $s=10$ and spacing 
$d$. The main difference arises from the behaviour at the resonance with the 
upper energy band. The phenomenology in high-$M$ lattices approaches the limit 
of infinite lattices, where gap solitons cannot live inside the bands. Saddle 
node bifurcations (dashed lines) appear in the second energy gap for $M>2$.}
 \label{fig:gap_bend}
\end{figure}
The allusion to bright 
solitons within the few-site lattice needs additional remarks, since the clear 
signature of a gap soliton, the localization, can only be manifested in a long 
lattice (compared with the spacing $d$). Once a bright soliton candidate has 
been found in a small lattice with  $M_0 > 1$, we look at equivalent 
states in longer lattices by keeping the spacing $d$ constant and 
increasing the number of sites $M$. In doing so the ring radius is 
increased, and the kinetic-energy quantum is reduced, so that the 
interaction parameter $\eta\propto R/M$ does not change. Additionally, to 
follow the transit towards the infinite lattice limit, we have to keep the 
parity in the number of sites so that the group velocity 
$\hat v=\partial_{q} E(q)$, and then the relative position of the state in the 
energy band, does 
not change. This constrains the lattice series to number 
of sites $M=(M_0)^j$, $j=1\,,2\,,3,\dots$ matching the powers of the original 
lattice $M_0$.

With this scheme, let us consider the family of gap solitons (born in the 
first energy gap) in a series of static ring lattices with $s=10$ and $M_0=2$, 
as shown in Fig.\ \ref{fig:gap_bend}, for $M=2,\,4,\,8,\,16$ and $M=64$. The 
gap-soliton trajectories in the $\mu$--$N$ graph and also their configurations 
are practically indistinguishable 
within the energy gap, a fact that allows us to study these states in low-$M$ 
lattices. Only the approach to the second 
energy band produces relevant differences. For increasing $M$ the 
trajectories bend towards the lowest energy state in the second band. The 
bending increase with $M$ approaching the limit of an infinite lattice, where 
gap solitons cannot live inside the energy band. However, the finite size of 
the rings releases this constraint. As soon as the gap solitons enter the 
bands, long oscillating tails are developed that are responsible of the 
observed bending in their trajectories. Notwithstanding, a clear density peak 
remains as the gap soliton signature. In what follows, we elaborate on these 
issues. 

\subsubsection{Origin at the non-interacting regime}

\begin{figure}[t]
 \centering
 \includegraphics[width=\linewidth]{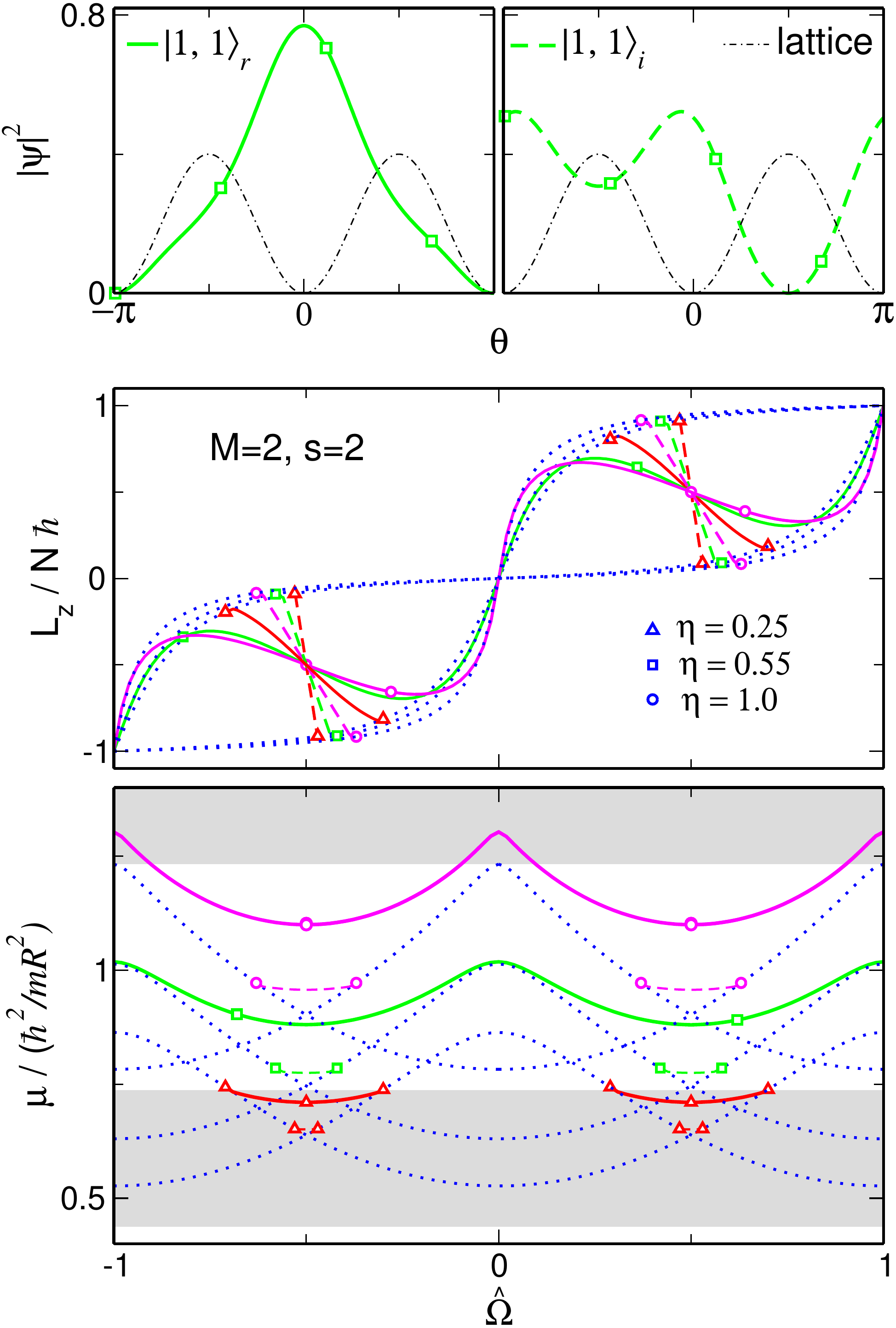}
 \caption{Origin and evolution of gap solitons in a two-well rotating ring 
lattice with $s=2$ for varying interaction strength $\eta$. The two 
lower panels show the mean angular momentum and the chemical potential of the 
soliton families (solid and dashed lines) originating in the first energy 
band, where they connect the Bloch-wave states $|1,0\rangle$ and $|1,1\rangle$ 
(dotted lines). The top panels depicts typical gap soliton density 
profiles (see text) at $\hat\Omega=0.5$ and  $\eta=0.55$.
}
 \label{fig:M2omega}
\end{figure}

As we have anticipated, gap solitons originate from linear degeneracies at the 
linear regime. There, depending on the parity of the number of lattice sites 
$M$, the bifurcation takes place either at $\hat\Omega\neq0$, when $M$ is even, 
or at $\hat\Omega=0$, when $M$ is odd. 

For even $M$ there is just one (hence 
non degenerate) linear Bloch state at the edge of Brillouin zone (with 
$q=M/2$). In this case, if one keeps $\hat\Omega=0$ the gap soliton bifurcation 
is 
observed to emerge from the highest energy states of the  
nonlinear energy bands (see Figs. \ref{fig:M34nonlinear} and \ref{fig:gs_types} 
for $M=4$), as usually reported. However, the gap soliton families originates 
before, within the linear regime for half integer values of $\hat\Omega$, just 
at the linear superposition of degenerate Bloch waves. In particular, the 
fundamental gap solitons (localized in a single site, like states A and B in 
Fig.\ \ref{fig:gs_types}) arise from the highest energy degeneracy in the 
energy bands. The observed nonlinear bifurcation at $\hat\Omega=0$ occurs when 
the whole family of gap solitons (forming a continuous energy band) detaches 
from the Bloch waves that were originally connected. 

The described process is summarized in Fig.\ \ref{fig:M2omega} for a 
minimal system with $M=2$ and $s=2$. The solid (dashed) lines correspond to a
solitonic family originated from the real 
(complex) linear combination of degenerate Bloch waves at $\hat\Omega=0$: 
$|1,1\rangle_r=|1,1\rangle + |1,0\rangle$ and $|1,1\rangle_i=|1,1\rangle +i\, 
|1,0\rangle$. These families, represented at different 
interactions values in the two lower panels, present one and 
two density peaks in the ring, respectively (top panel of Fig.\ 
\ref{fig:M2omega}). Their trajectories in the $\mu$--$\Omega$ graph shape 
swallow tail structures, that grow with the interaction, in the interior of 
the energy bands. They are not different from the intra-band swallowtails  
previously associated at dark-soliton-like states.
At $\eta\approx0.55$, the family of fundamental solitons (solid green curve) 
detaches inside the energy gap from the connected Bloch waves. An equivalent 
process 
takes place for gap solitons emerging in higher energy gaps.
 
This phenomenon is equivalent to the metastable quantum phase transition 
described in Ref. \cite{Kanamoto2009} for states with attractive interaction in 
a ring without lattice. There the bright soliton solutions separate from the 
winding number states at a given interaction threshold, and the solitons 
always have lower energy than the winding number states for given $\hat\Omega$. 
On the contrary, the localized states in the lattice have higher energy than 
the Bloch states, and, importantly, the curve separation (the bifurcation 
at $\hat\Omega=0$) occurs inside the linear energy gap. The interaction value at
separation varies with the radius of the ring and the depth of the lattice. Our 
numerical results show that the bifurcation point gets closer to 
the linear limit as the system approaches to the infinite lattice 
configuration. This fact can be qualitatively understood as a result of the 
changes produced in the energy bands, featured by a higher number of available 
states that are separated by lower energy increments. Under these conditions, 
the system can 
transfer such differences in kinetic energy of the extended states into 
interaction energy of a localized state at essentially the same value of the 
chemical potential. An equivalent mechanism operates in deeper lattices, where 
there also exist a higher density of available states because of the reduced 
band widths. 

For odd $M$ there are two degenerate stationary states with definite 
quasimomentum at $|q_{M}|<M/2$, the highest absolute value of the discrete 
band.  In the linear regime, at $\hat\Omega=0$, there are two independent 
linear combinations $\alpha|q_{M}\rangle+\beta|-q_{M}\rangle$ that show 
the typical features of gap solitons. Figures \ref{fig:M34nonlinear} (top 
panel),  \ref{fig:M34omega}, and \ref{fig:M34wf} illustrate the origin and 
evolution of gap solitons in a lattice with $M=3$. The symmetric states, 
continuation of 
$|1,1\rangle_s=|1,1\rangle + |1,-1\rangle$, become more energetic 
than the antisymmetric ones, continuation of $|1,1\rangle_a=|1,1\rangle - 
|1,-1\rangle$). Both families are made of real stationary states with zero mean 
angular momentum, and unlike the states with definite $q$,  present  
inhomogeneous density profiles that break the symmetry of the lattice. 
The gap soliton curves are already detached from the Bloch wave states. 

The differences introduced by the parity of $M$ in the few-site ring lattice 
are strongly reduced in long (high $M$) lattices. In the infinite lattice limit 
the gap soliton family bifurcates from the linear regime at $\hat\Omega=0$,  
although near the bifurcation point the states belonging to this branch can not 
show a sharp localization. The longer the lattice the more advance within the 
nonlinear regime is needed in order to get a clear-cut density peak.

\subsubsection{Resonances with the linear energy bands}
\begin{figure}[tb]
 \centering
 \includegraphics[width=0.95\linewidth]{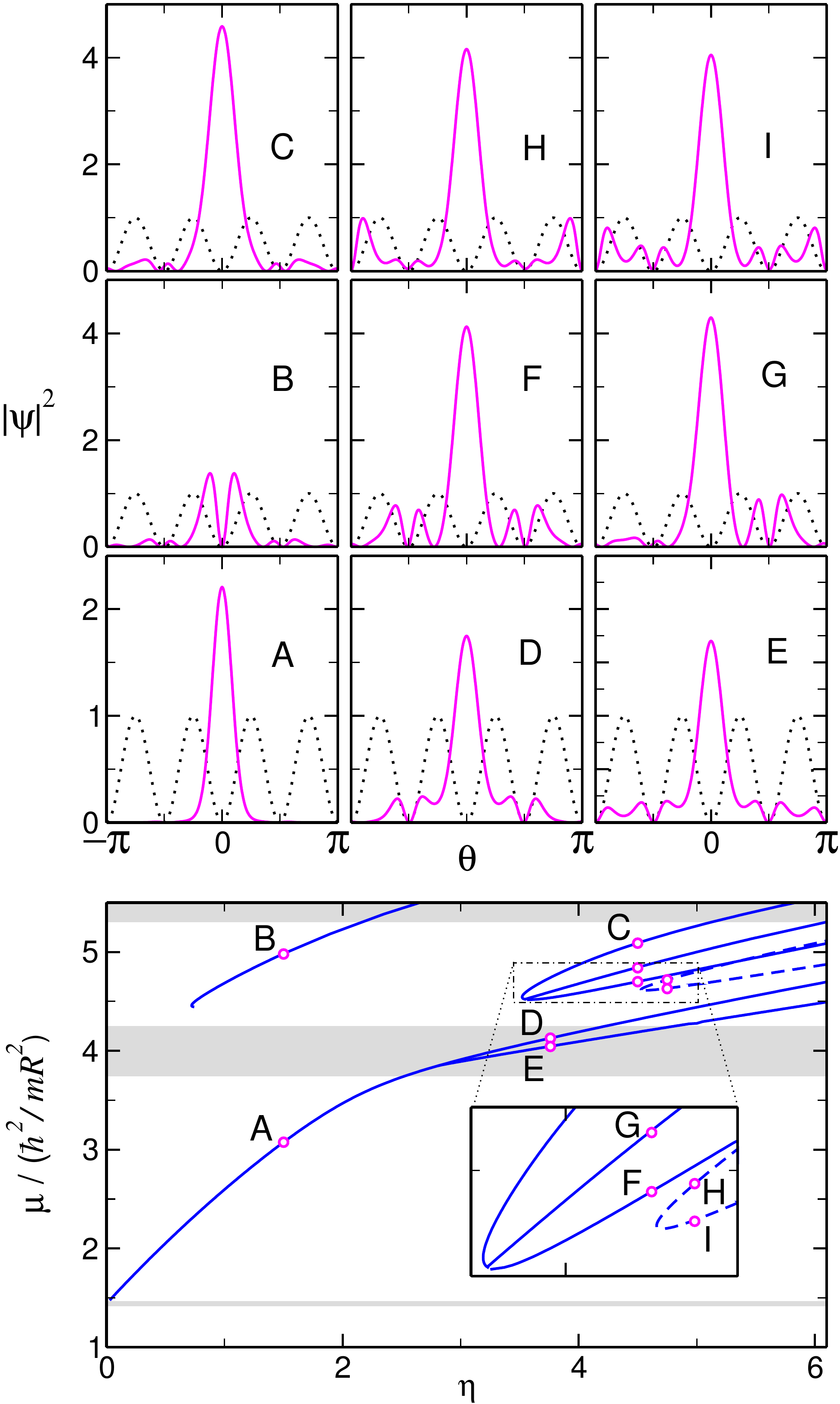}
 \caption{(a) Gap solitons in a ring lattice with $M=4$,  
$s=10$, and $\hat\Omega=0$. The bottom panel shows the trajectories of gap 
solitons in the  $\mu$-$\eta$ graph over the underlying 
linear energy bands (grey-shadowed stripes). The upper panels labeled A--I 
depict the number density of the states 
(solid lines) marked by open symbols in the bottom panel; for reference, 
the lattice is represented by dotted lines.}
\label{fig:gs_types}
\end{figure}
We have summarized the configurations and bifurcations of bright soliton states 
in Fig.\ \ref{fig:gs_types}, inside a static ring lattice with $M=4$ sites and 
depth $s=10$. Within each energy gap a new family of bright solitons 
appears \cite{Zhang2009}; this is the case of solitons A and B inside the first 
and the second gap respectively. These solitons present $n-1$ nodes ($n$ being 
the lower band or gap index) inside the lattice site where they are localized. 
As we show below, these states can be traced back to the linear regime, where 
they originate in the interior of the energy band immediately below the 
gap where they first emerge. The other types of gap solitons (C to I in Fig.\ 
\ref{fig:gs_types}) bifurcate from resonances with 
excitation modes proceeding from the linear Bloch waves. Among the 
bifurcations, the saddle node bifurcation is particularly relevant since it is 
responsible for the continuation of the soliton trajectories into upper energy 
gaps (like soliton C).

\begin{figure}[tb]
 \centering
 \includegraphics[width=\linewidth]{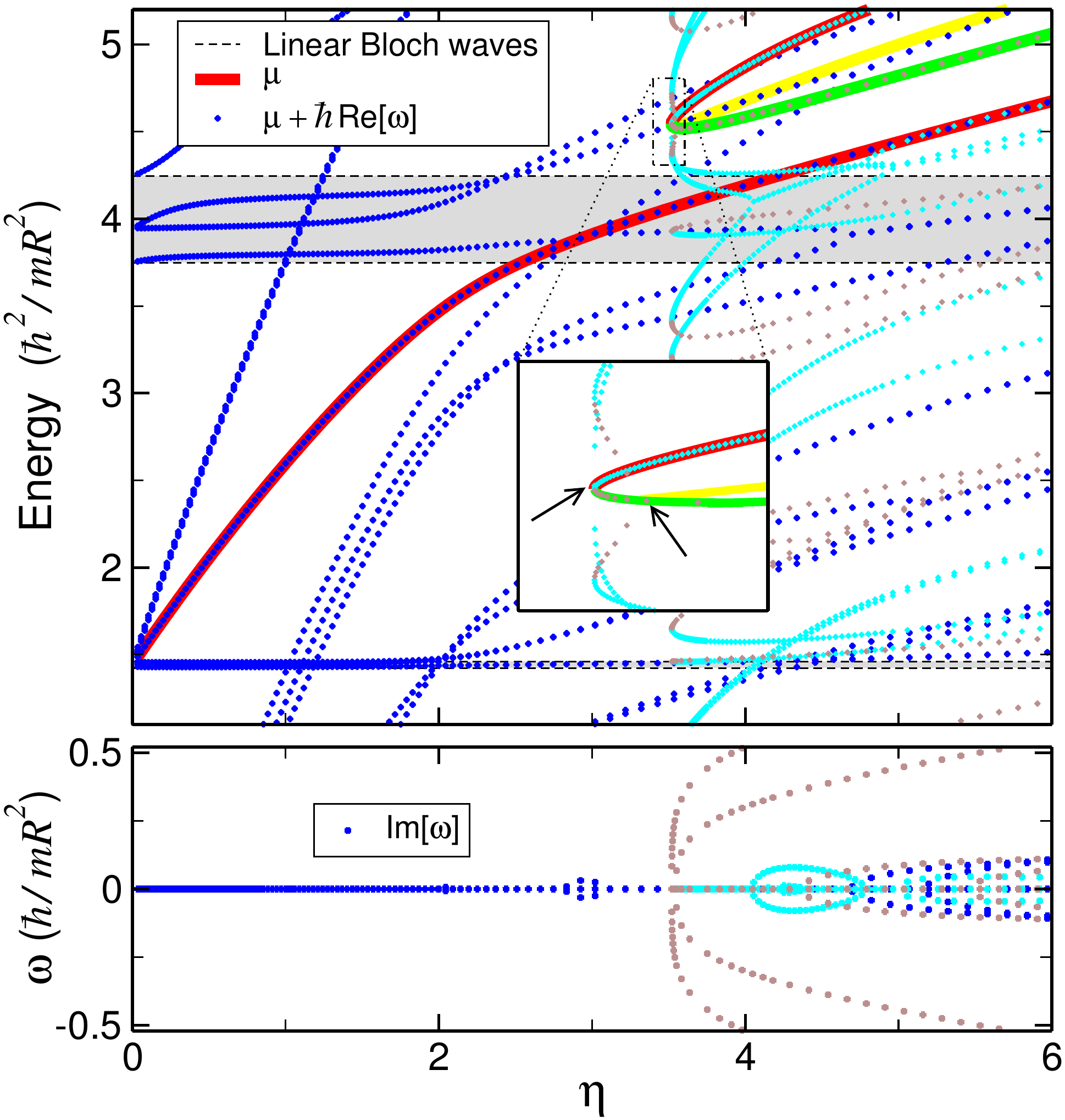}
 \caption{ Top panel: Chemical potential of
fundamental gap solitons (thick lines) in the ring lattice of Fig.\ 
\ref{fig:gs_types}, and, superimposed, sum of chemical
potential and real part of Bogoliubov excitation energies
(symbols) as a function of the interaction. The arrows in the inset 
point to the saddle node bifurcation (left arrow), due to the collision of gap 
solitons of type C and F in Fig.\ \ref{fig:gs_types}, and to a secondary 
bifurcation (right arrow) of solitons F to give rise to the family of states G.
Bottom panel: Imaginary part of the frequencies of Bogoliubov excitations
as a function of interaction.
}
 \label{fig:M4saddle}
\end{figure}
The resonance of the bright soliton frequency with those of the linear 
spectrum is the key for the non-existence of localized states inside the 
energy bands of infinite lattices \cite{Delgado2018}. In this regard, plausible 
arguments have 
been given in the realm of discrete systems \cite{Flach1998}. Such resonances 
produce oscillating tails surrounding the soliton that have the structure of 
the resonant linear Bloch waves. The extended tails eventually dismantle the 
localized density peak. Practically the same happens in the ring 
lattice, with the peculiarity of the finite system to admit the coexistence of 
high density peaks on the extended density tail. 

Differently from the case of out-of-band swallow tails, gap solitons present 
linear excitations (proceeding from the top of the energy bands) that diverge 
from the underlying linear spectrum, along with excitation energies that keep 
a roughly constant value close (the longer the lattice the closer) to the 
linear spectrum (see Fig.\ \ref{fig:M4saddle}). The collisions between constant 
and diverging excitation modes create a complex scenario of instability 
regions. Figure \ref{fig:M4saddle} illustrates this behavior for gap solitons 
in the static lattice with $M=4$ and $s=10$ of Fig.\ \ref{fig:gs_types}. 
Before (but close to) entering the energy band, the gap solitons in the first 
gap present instabilities (in small ranges of $\eta$) triggered by the 
collision at non zero frequency $\hat \omega$ between the 
modes proceeding from the first and the third energy bands.
When the gap solitons enter the energy band (like state D in Fig.\ 
\ref{fig:gs_types}), the appearance of an imaginary excitation frequency 
marks a pitchfork bifurcation of new  families of soliton-like states (made of
current-carrying states like E). 

Within the second gap, a saddle node bifurcation is 
produced by the collision of two types of gap soliton solutions 
with different symmetry ( C and F in Fig.\ 
\ref{fig:gs_types}). For this bifurcation to take place, 
it is necessary that $M>2$, since excitation modes with intermediate 
quasimomentum proceeding from the linear energy bands  (other than those 
proceeding from the Brillouin zone edge) are needed to provide the specific 
density pattern in the neighbor sites of the soliton. Such modes are also 
allowing for alternative density patterns in the soliton tail that produce 
secondary bifurcations (as the branch containing state G). Furthermore, 
additional saddle node bifurcations involving 
current states (like H and I) take place at higher interaction. For incresing 
$M$, there is also an increasing number of  density configurations available
around the soliton peaks, and therefore longer lattices entail larger sets 
of solitonic states.

\subsubsection{Gap solitons in longer lattices}

\begin{figure}[t]
 \centering
 \includegraphics[width=\linewidth]{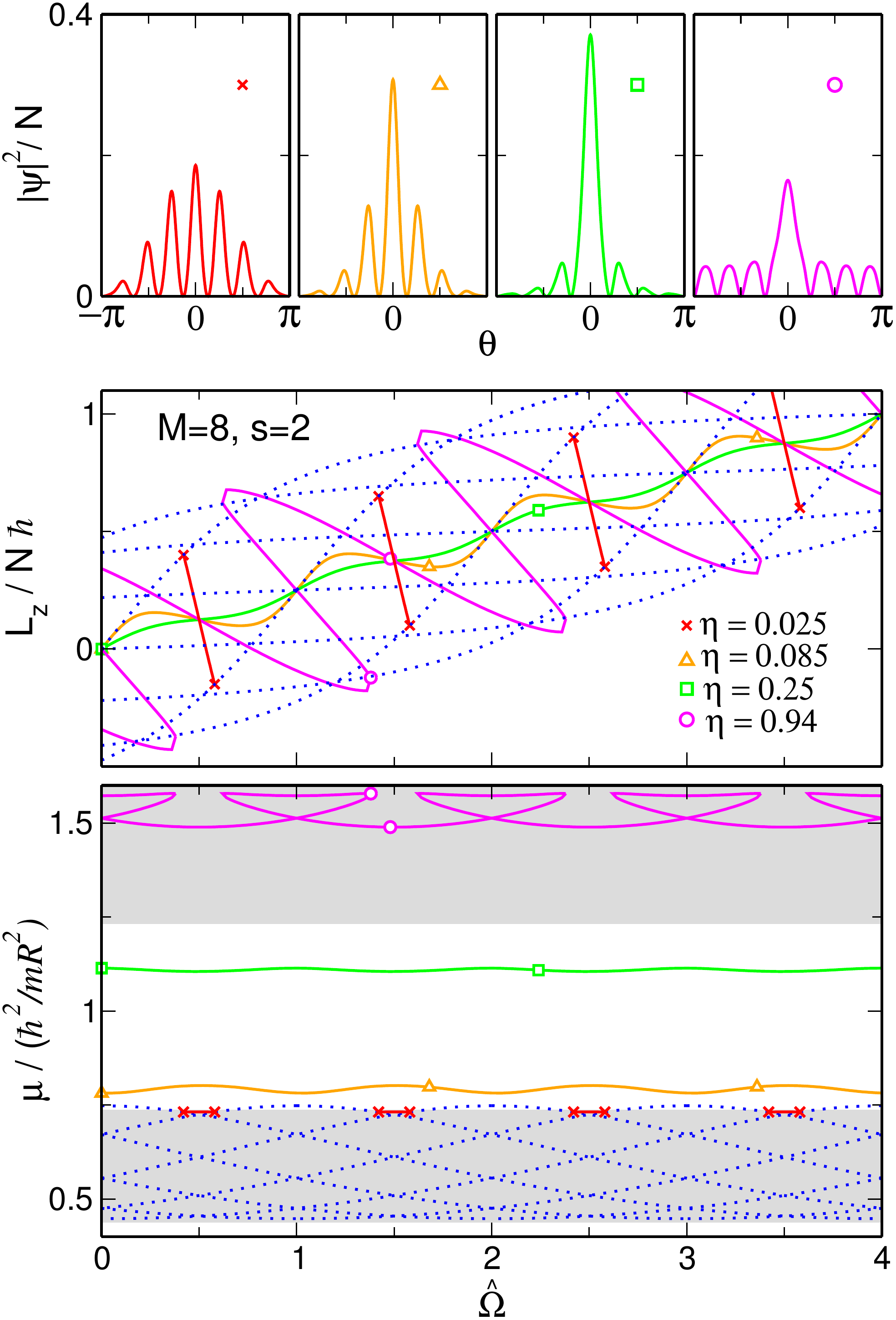}
 \caption{Gap soliton generation in a 8-site ring lattice with $s=2$ and
varying interaction parameter $\eta=\hat g N/M$.
The panels in the top row depict the soliton density at $\hat\Omega=0.5$ for 
increasing $\eta$. The two lower panels show the soliton angular momentum and 
chemical potential for varying rotation rate $\hat\Omega$. The dotted lines 
represent the nonlinear Bloch waves.
}
 \label{fig:M8omega}
\end{figure}
In order to see the correspondence with longer lattices, we have chosen a 
system with $M=8$ and equal lattice spacing as the minimal system of Fig.\ 
\ref{fig:M2omega}. In Fig.\ \ref{fig:M8omega}, one can see that the mechanism 
for the generation of bright solitons is the same. They emerge from the linear 
regime and from the connection between Bloch waves. The separation of the 
chemical potential curve takes place within the energy gaps for interaction 
energies that are lower than those found in the shorter lattice. 
In addition, the soliton energy band becomes flatter 
far from the first (linear) energy band, showing particle-like features, i.e. 
flat energy in the co-moving reference frame, and linear increase in momentum. 
For higher nonlinearity,  the resonance with the second energy band 
increases the soliton band width and leads to a bifurcation with the 
generation of swallow tails similar to those developed by nonlinear Bloch waves.

\section{Conclusions}

The small ring lattices are minimal systems that can be mapped to the infinite 
lattice 
by the introduction of angular rotation. In this way, and due to the existence 
of a small number of available states, the ring lattices allow for a simpler 
analysis of the states of equilibrium. In particular,
we have addressed the origin, range of existence, and stability of solitonic 
states. Our results show that they share a general common 
origin with the nonlinear Bloch waves that can be tracked up to the linear 
regime. While the Bloch waves are states with definite quasimomentum that 
follow the symmetry of the underlying lattice, the solitonic states 
emerge as linear combination of Bloch waves and break this symmetry.
The symmetry-breaking states have both a nodal structure as 
well as a local density maximum and could thus be classified as either dark or 
bright solitons.
Clear-cut features are only shown in the dark-soliton states of the 
out-of-band swallow tails, or in the (fundamental) gap-soliton states that 
bifurcate from the top of the energy bands.

The resonances of the nonlinear waves with the underlying linear 
spectrum produce dynamical instabilities and new bifurcations. Among them, we 
have shown how the pitchfork bifurcation of the highest energy nonlinear Bloch 
waves give rise to out-of-band swallow tails, how the saddle node bifurcation 
of 
fundamental gap solitons continue the soliton families into higher energy gaps, 
and how the ring lattice provides a plausible demonstration of the nonexistence 
of localized states inside the linear energy bands of the infinite lattice. 

The small ring lattices could play an interesting role in BEC experiments in 
order to facilitate the materialization of controlled persistent currents or 
even solitonic states. Previous experiments have made use of a 
Gaussian-shaped barrier potential
to drag a toroidal system into a persistent current state 
\cite{Eckel2014}, but this procedure involves sudden phase slips that produce 
a transitory unstable stage. The present work shows that a periodic 
potential can provide an simpler and adiabatic way to transit between quantized 
persistent currents. The realization of 
solitonic states seems  more challenging, however, because the lattice 
symmetry has to be broken. This could be achieved
by the combination of a precise rotation 
rate and a symmetry-broken initial state in analogy to the experiment of
Ref.~\cite{Eiermann2004} where gap solitons were produced in a linear 
lattice.

\begin{acknowledgments}
A.\ M.\ M.\ is grateful to Sergej Flach for insightful discussions.
V.\ D.\ acknowledges financial support from Ministerio de Econom{\'i}a y 
Competitividad (Spain) and  Fondo Europeo de Desarrollo Regional (FEDER, EU) 
under Grants No.\ FIS2013-41532-P and FIS2016-79596-P.
M.\ G.\ and R.\ M.\ acknowledge financial support from Ministerio de Econom{\'i}a y 
Competitividad (Spain) and Fondo Europeo de Desarrollo Regional (FEDER, EU) 
under Grants No.\ FIS2014-52285-C2-1-P and FIS2017-87801-P. 
J.\ B.\ acknowledges funding by the Marsden Fund of New Zealand
(contract no.\ MAU1604), from government funding managed by the
Royal Society Te Ap\=arangi.
\end{acknowledgments}

\bibliography{ring_lattice_0}

\end{document}